\documentclass{aa}
\pdfoutput=1
\usepackage[version=3]{mhchem} 
\usepackage{graphicx} 
\usepackage[nice]{nicefrac} 
\usepackage{multirow} 
\usepackage{url}
\usepackage{nccmath} 
\usepackage{booktabs}

\begin{document}

\title{Modeling Sulfur Depletion in Interstellar Clouds}

\author{Jacob C. Laas\thanks{jclaas@mpe.mpg.de} \and Paola Caselli\thanks{caselli@mpe.mpg.de}}
\institute{Max-Planck-Institut f\"ur extraterrestrische Physik, Garching, DE 85748}

\titlerunning{Modeling Interstellar Sulfur Depletion}
\authorrunning{Laas \& Caselli}


\abstract
{The elemental depletion of interstellar sulfur from the gas phase has been a recurring challenge for astrochemical models. Observations show that sulfur remains relatively non-depleted with respect to its cosmic value throughout the diffuse and translucent stages of an interstellar molecular cloud, but its atomic and molecular gas-phase constituents cannot account for this cosmic value towards lines of sight containing higher-density environments.}
{We have attempted to address this issue by modeling the evolution of an interstellar cloud from its pristine state as a diffuse atomic cloud to a molecular environment of much higher density, using a gas/grain astrochemical code and an enhanced sulfur reaction network.}
{A common gas/grain astrochemical reaction network has been systematically updated and greatly extended based on previous literature and previous sulfur models, with a focus on the grain chemistry and processes. A simple astrochemical model was used to benchmark the resulting network updates, and the results of the model were compared to typical astronomical observations sourced from the literature.}
{Our new gas/grain astrochemical model is able to reproduce the elemental depletion of sulfur, whereby sulfur can be depleted from the gas-phase by two orders of magnitude, and that this process may occur under dark cloud conditions if the cloud has a chemical age of at least $10^6$ years. The resulting mix of sulfur-bearing species on the grain ranges across all the most common chemical elements (H/C/N/O), not dissimilar to the molecules observed in cometary environments. Notably, this mixture is not dominated simply by \ce{H2S}, unlike all other current astrochemical models.}
{Despite our relatively simple physical model, most of the known gas-phase S-bearing molecular abundances are accurately reproduced under dense conditions, however they are not expected to be the primary molecular sinks of sulfur. Our model predicts that most of the ``missing'' sulfur is in the form of organo-sulfur species trapped on grains.}

\keywords{astrochemistry -- molecular processes -- ISM: molecules}

\maketitle


\section{Introduction}

Sulfur poses an interesting challenge to models of interstellar chemistry. Within primitive interstellar environments, it is known that sulfur remains in ionized atomic form and close to the cosmic abundance \citep{Jenkins2009}. However, in molecular clouds and star-forming regions, this cosmic abundance is drastically reduced from the gas-phase molecular inventory. Many simple organo-sulfur species can be detected, but not seemingly enough to fully account for its cosmic abundances.

Modern astrochemical models still today predict that the bulk of sulfur resides as condensed \ce{H2S}, despite upper limits from observations \citep{Smith1991, vanderTak2003, JE2011}. The effective workaround has been to use a severely depleted value of the elemental abundance ($\lesssim 1\%$ with respect to the cosmic standard abundance) for the initial sulfur content when modeling dense interstellar environments. This has resulted in a rather poor understanding of sulfur with respect to interstellar matter, and sulfur remains a severe shortcoming of astrochemical models. That's not to say that astrochemical models of sulfur are not improving. For example, a recent model \citep{Woods2015} has helped to provide constraints on how much sulfur may be locked in the refractory residue that is known from laboratory studies of photo-processed sulfur; the reaction network pertaining to certain categories of molecules has also been expanded by the study reported by \citet{Vidal2017}; and, a better understanding of dynamical models is provided by \citet{Vidal2018} to help put into perspective the apparent chemical variability of sulfur chemistry across cloud evolution.

Laboratory studies are also continuously helping to shed light on sulfur chemistry, which can be notoriously complex. Sulfur bonds are generally not as strong as those of the first- and second-row elements (i.e. H, C, N, O), and processing of interstellar analogs of ice mixtures containing simple S-bearing molecules can yield a highly heterogeneous mixture of products \citep[e.g.][and references therein]{JE2014}. Qualitatively, these processed ices sometimes even resemble the type of chemistry that has been detected in both cometary ices \citep{Calmonte2016} and meteoritic material \citep[see, e.g.][]{Ehrenfreund2002}. However, many important laboratory studies have not yet been incorporated into a modern gas/grain astrochemical model.

We have set out to update the sulfur chemistry within a modern astrochemical reaction network in a systematic and more complete way than has been done in the past, and then use this gas/grain reaction network to model the evolution of an interstellar cloud from its pristine diffuse stage to a dense and dark quiescent state to check for clues pertaining to the gas-phase depletion of sulfur. A collection of observational studies has also been used as general constraints for benchmarking this updated astrochemical model to ensure reasonable results for all the sulfur-bearing interstellar molecules that are typically observed in interstellar environments. We have found that this updated sulfur network does in fact reproduce observations of interstellar sulfur chemistry in dense environments, and this relatively simple astrochemical model also reproduces the severe gas-phase depletion of sulfur thanks to the production a variety of (mostly-stable) organo-sulfur molecules on grains (see \S\ref{sec:results}).

In \S\ref{sec:model}, we first present the astrochemical model in terms of its chemical and physical characteristics. In \S\ref{sec:results}, we present our results, beginning with a look at the question of elemental depletion of sulfur during the dense stage (m$_\text{gas}$ $\approx 10^4$--$10^6$ cm$^{-3}$) of an interstellar cloud, and then followed up with a brief analysis of commonly observed molecules across a range of evolutionary stages/densities of interstellar clouds with our model predictions. In \S\ref{sec:discussion} we provide a summary comparing our model with previous studies and to discuss key results. We finally end the manuscript with \S\ref{sec:conclusions} containing some general conclusions.


\section{Astrochemical model} \label{sec:model}


\subsection{Computer code}

The various model trials presented here have been performed using a new python-based code, which has not yet been described elsewhere. The code is based on the OSU gas/grain astrochemical model presented by \citet{Garrod2008}, which is a time-dependent, single-point, rate-based model for general application to interstellar cloud environments. Benchmarks were performed during its development to ensure that its operation is consistent with the code it is based on, and it has been further enhanced with a number of internal consistency checks and software tools for the development of a chemical network and for exploring its results.


\subsection{Chemical network}

The core reaction network was based on the OSU gas/grain model, which consists of ca. 8300 reactions and 860 chemical species, and remains in high circulation even a decade after being released. Aside from a major revamping of the sulfur chemistry---which we address in a separate section below---a number of modifications were made to the chemical inputs to reflect recent literature reports. These changes include: a) modifications to the thermodynamical properties of many grain surface species; b) updated photochemical rates as per the report from \citet{Heays2017}; and, c) a number of other updates to various reactions. For completion, all the binding energies and heats of formation for the grain species are listed in Table \ref{table:binde}; except for all the updates from \citet{Heays2017}, we also list out new/modified reactions in Table \ref{table:nonSchanges}. In the particular, we have added routes for the heavy atoms (\ce{Fe+}, \ce{Mg+}, \ce{Na+}, \ce{S+}, \ce{Si+}, \ce{Cl+}, and \ce{P+}) to adsorb directly onto negatively-charged grains, as discussed in \citet{Ruffle1999} and references therein.

Whereas a number of photodesorption reactions have been added based on \citet{Hollenbach2009}, we have completely disabled/ignored chemical (i.e. reactive) desorption. The new photodesorption entries are crucial for accurate results during Stages 1 and 2 (i.e. the diffuse and translucent cloud phases), however, recent studies suggest that chemical desorption is largely not efficient in ices \citep{Minissale2016, Chuang2018} and additionally introduces an added complexity which we felt detracts too much focus from development on the chemical network and the study of sulfur depletion.

We have used, when possible, known cosmic standard elemental abundances for the initial abundances of our model (i.e. for Stage 1), and these values are contained in Table \ref{table:initabunds}. Whereas the details of the stages are presented below, in \S\ref{sec:physics}, we note here that the initial fractional abundances of subsequent stages are simply the final abundances of the preceding stages.

\begin{table}\centering
\caption{Initial Elemental Abundances}
\label{table:initabunds}
\begin{tabular}{lrr}
\hline
\hline
Species & Fractional~~ & Reference \\
        & Abundance\tablefootmark{a} & \\
\hline
\ce{H} & 0.9999 &  \\
\ce{H2} & 5e$-$5 &  \\
\ce{He} & 9.55e$-$2 & 1 \\
\ce{O} & 5.7544e$-$4 & 1 \\
\ce{C+} & 2.0893e$-$4 &1 \\
\ce{N} & 5.7544e$-$5 & 1 \\
\ce{Mg+} & 3.6308e$-$5 & 1 \\
\ce{Si+} & 3.1623e$-$5 & 1 \\
\ce{Fe+} & 2.7542e$-$5 & 1 \\
\ce{S+} & 1.66e$-$5 & 2 \\
\ce{Na+} & 1.74e$-$6 & 3 \\
\ce{Cl+} & 2.88e$-$7 & 2 \\
\ce{P+} & 2.57e$-$7 & 3 \\
\ce{F} & 3.63e$-$8 & 3 \\
\ce{e-} & 3.233e$-$4 & \tablefootmark{b} \\
\hline
\end{tabular}
\tablefoot{
\tablefoottext{a}{We define the fractional abundance $X(i)$ as the ratio of the abundance of species $i$ wrt the total gas-phase hydrogen number density, which is initially equal to $X(\ce{H}) + 2X(\ce{H2})$.}
\tablefoottext{b}{The total electron fraction is simply initialized as the sum of the gas-phase cations.}}
\tablebib{
(1)~\citet{Przybilla2008};
(2)~\citet{Esteban2004};
(3)~\citet{Asplund2009}.
}
\end{table}

\subsubsection{Sulfur chemistry}

In order to allow for a number of chemical routes that a high (cosmic) sulfur abundance may take, we have sorted through a wealth of literature data to systematically build a more complete gas/grain sulfur network. These new reactions were based on both experimental studies relating specifically to interstellar ice analogs and also to atmospheric chemistry. Furthermore, sulfur is known for its rich chemistry, and behaves more similarly to carbon than to oxygen, despite being found below the latter in the periodic table of the elements. The sulfur chemistry in our network takes into account oxidation states and a plurality of isomers and bond types.

As with the other heavy atoms and what we have pointed out above, our model includes a route for direct accretion of \ce{S+} onto negatively-charged grains. To counterbalance this effect, we found that it was important to add also the photodesorption of \ce{H2S}, so that it behaves similarly to a number of other simple species---in particular, \ce{CH4}, \ce{CO} and \ce{H2O}.

In Tables \ref{table:removedrxns}--\ref{table:newrxns}, we list all of the new reactions which have supplemented the base network. In general, many of the included rates have been estimated from scratch, based on similar chemistry (yielding reasonably large uncertainties) or calculated from basic electrostatics data and capture rate theory \citep{Troe1985}. We also provide highly-detailed explanations/notes for these collections of reactions, which can be found also within the Appendix.


\subsection{Physical model}\label{sec:physics}

The physical model we have employed is defined by three distinct, physically-static stages. Whereas the code is capable of time-dependent densities and temperatures, we have opted to use only static physics for each stage, to avoid significant uncertainties related to the dynamical evolution of molecular clouds. Here we focus on the effects of the new chemistry for comparison with previous models, which are also mainly static; a dynamical-chemical model is deferred to future work.

Stage 1 follows the evolution of a diffuse cloud during its atomic-to-molecular phase transition, with the final time of this transition being defined as the moment when less than 50\% of the original atomic hydrogen content remains. Stage 2 simulates a translucent cloud, whose defining characteristics are its increased visual extinction compared to diffuse clouds, as well as the conversion of ionized carbon to molecular species. Lastly, Stage 3 is meant to simulate the aging of a dense core, and we have included a range of densities in order to explore the effect of density on chemical differentiation and timescales, as this is known to be a varying parameter across astrophysical environments and should play a critical role on the kinetics/rate of sulfur depletion. We employ the convention of resetting the defined time to $t=0$ at the beginning of each stage, and the initial abundances of Stages 2 and 3 are set to those which have been predicted in the preceding stage at the times defined in Table \ref{table:cloudstages}.

These three stages adhere well to the categorization of interstellar clouds presented by \citet[][\S2]{SnowMcCall2006}, and their defining characteristics are summarized in Table \ref{table:cloudstages}. We also show below (Figure \ref{fig:HCOe_early}) that this sequence of stages is able to approximate well the main astrochemical results supported by observations during the early stages of molecular cloud evolution, agreeing particularly well with Figure 1 from \citet{SnowMcCall2006}.

\begin{table}\centering
\caption{Three-stage Evolutionary Cloud Model}
\label{table:cloudstages}
\begin{tabular}{llll}
\hline
\hline
 & Stage 1 & Stage 2 & Stage 3 \\
\hline
Classification\tablefootmark{a}         & Diffuse         & Translucent & Dense \\
n$_\text{gas}$ (cm$^{-3}$)     & 1e2             & 1e3         & $10^4$--$10^6$ \\
N(\ce{H2})$_\text{ext.}$ (cm$^{-2}$)\tablefootmark{b} & 5.7e20          & 5.7e20      & 3.1e21 \\
radius (pc)                    & 25              & 10          & 1 \\
A$_\text{V}$ (mag)             & 0.5             & 1.6         & 10 \\
$\zeta$ (s$^{-1}$)             & 1.0e-16\tablefootmark{c} & 1.3e-17     & 1.3e-17 \\
T$_\text{gas}$ (K)             & 100             & 25          & 10 \\
T$_\text{dust}$ (K)            & 15              & 15          & 10 \\
Time (yr)\tablefootmark{d}              & $10^7$          & $10^6$      & $10^5$--$10^6$ \\
\hline
\end{tabular}
\tablefoot{
\tablefoottext{a}{Classification is based on \citet{SnowMcCall2006}; see text for details.}
\tablefoottext{b}{This the \ce{H2} column density external to our single-point calculations, whereby the total column density used for the shielding factor is the sum of the external and ``internal'' components; the component of external N(\ce{CO}) is always set to a factor of $10^{-6}$ below that of the external N(\ce{H2}), as per general agreement with observations \citep[][and references therein]{SnowMcCall2006}.}
\tablefoottext{c}{Equivalent to that of Rho Oph (Indriolo+ 2012).}
\tablefoottext{d}{These times denote the final time that we allow for the model and what we use to compare to observations.}}
\end{table}

Whereas the model is a single-point calculation, the column densities and radii listed above are used for determining the total effective column densities of \ce{H2} and \ce{CO}, which are then used to determine on-the-fly their self-shielding factors for photodissociation. For the work presented here, we have used parameterizations based on \citet[][Eqn. 37]{Draine1996} and \citet{Lee1996}, for \ce{H2} and \ce{CO}, respectively.

Besides the time-/stage-dependent physical parameters listed above, a large number of other parameters are involved behind the scenes, particularly for reactions on the grain. These additional parameters are standard values nowadays, but for completeness' sake we list them in Table \ref{table:miscparams}.

\begin{table}
\caption{Other Adopted Parameters}
\label{table:miscparams}
\begin{tabular}{lr}
\hline
\hline
Input Parameter & Value \\
\hline
dust-to-gas mass ratio & 0.01 \\
dust mass density & 3 g cm$^{-3}$ \\
dust radius & 0.1 $\mu$m \\
CR-induced grain heating peak temp & 70 K \\
CR-induced grain heating duration & 1e-5 s \\
CR-grain encounter rate & 3e-14 s$^{-1}$ grain$^{-1}$ \\
FUV flux rate & 3.38e-8 s$^{-1}$ cm$^{-2}$ \\
\nicefrac{E$_\text{diff}$}{E$_\text{bind}$} & 0.4 \\
sticking probability & 1 \\
grain site spacing & 1 {\AA} \\
grain site density & 1.5e15 cm$^{-2}$ \\
grain sites & 1e6 grain$^{-1}$ \\
grain quantum barrier thickness & 1 {\AA} \\
\hline
\end{tabular}
\end{table}


\section{Results} \label{sec:results}


\subsection{General early-Stage features}

\begin{figure*}
\resizebox{\hsize}{!}{\includegraphics[trim={0.15cm 0.1cm 0.35cm 0.3cm},clip]{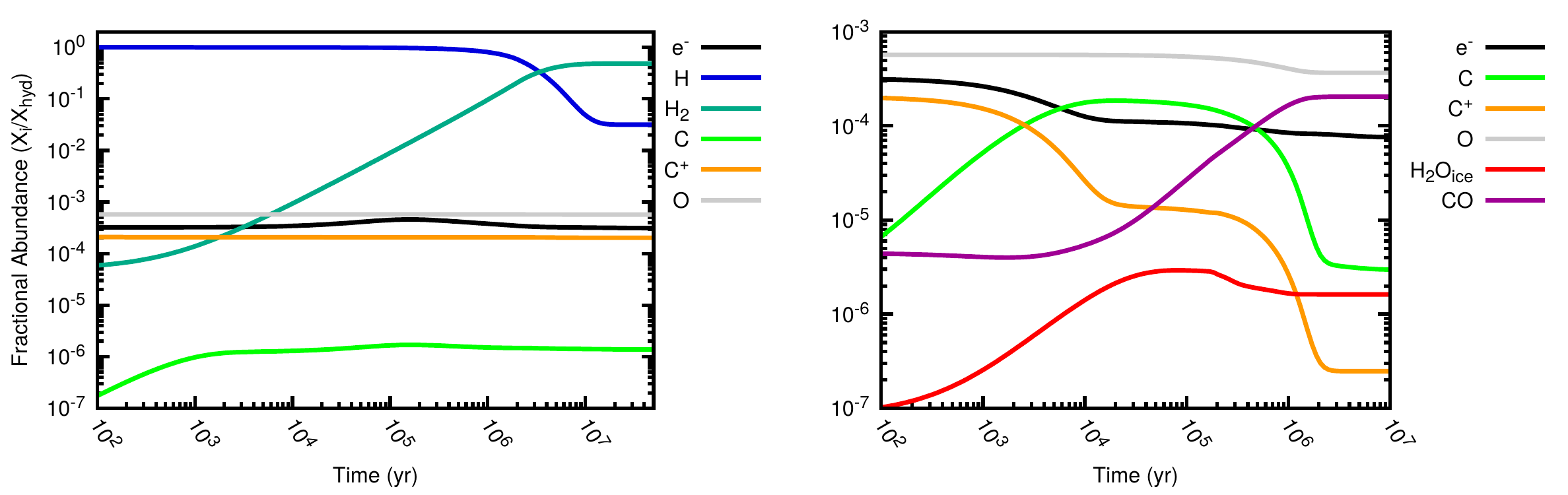}} 
\caption{Fractional abundances of the H/C/O elements during their evolution across the diffuse (\textit{left}) and translucent (\textit{right}) stages.\label{fig:HCOe_early}}
\end{figure*}

In the plots shown in Figure \ref{fig:HCOe_early}, we present the time evolution of key astrochemical species during the early stages of cloud evolution, showing consistent agreement with the defining chemical characteristics summarized by \citet[][\S2, Figure 1]{SnowMcCall2006} for the diffuse-to-translucent phase transition of an interstellar cloud. The single significant exception with respect to \citet{SnowMcCall2006} is the ionization fraction, which does not deplete so quickly in our model, continuing to be supported by the heavy atoms \ce{Mg+}, \ce{Si+}, \ce{S+}, \& \ce{Fe+} after carbon has been significantly neutralized. This steadier ionization fraction is somewhat consistent with models of translucent clouds ($A_v \leq 2$ mag) \citep[e.g.][]{Herbst1986, Cecchi2000}. While none of these results are particularly novel, they clearly benchmark our gas/grain astrochemical model against a general consensus of astronomical observations, providing reasonable context for the comparisons of S-bearing species presented below.

\subsection{Sulfur depletion and its primary budget}

\begin{figure*}\centering
\resizebox{\hsize}{!}{\includegraphics[trim={0.15cm 0.1cm 0.1cm 0.3cm},clip]{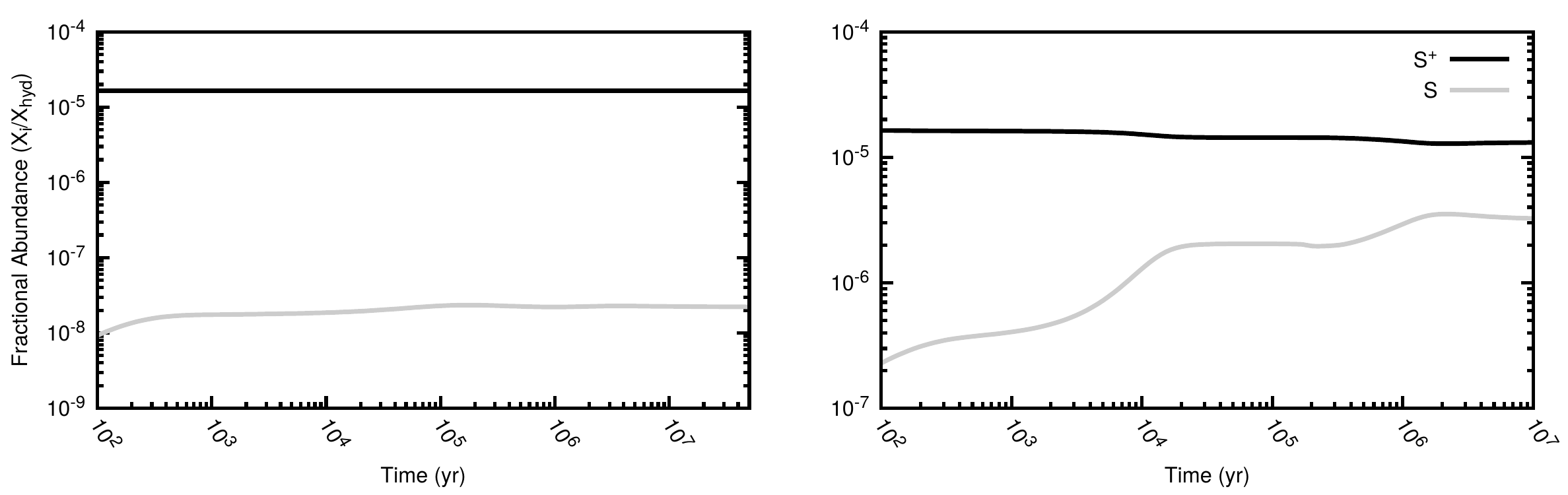}}
\caption{Time plots of atomic (neutral and singly-ionized) sulfur during the diffuse (\textit{left}) and translucent (\textit{right}) stages.\label{fig:sulfursinks_early}}
\end{figure*}

\begin{figure*}\centering
\resizebox{0.82\hsize}{!}{\includegraphics[trim={0 0.5cm 0 1.1cm},clip]{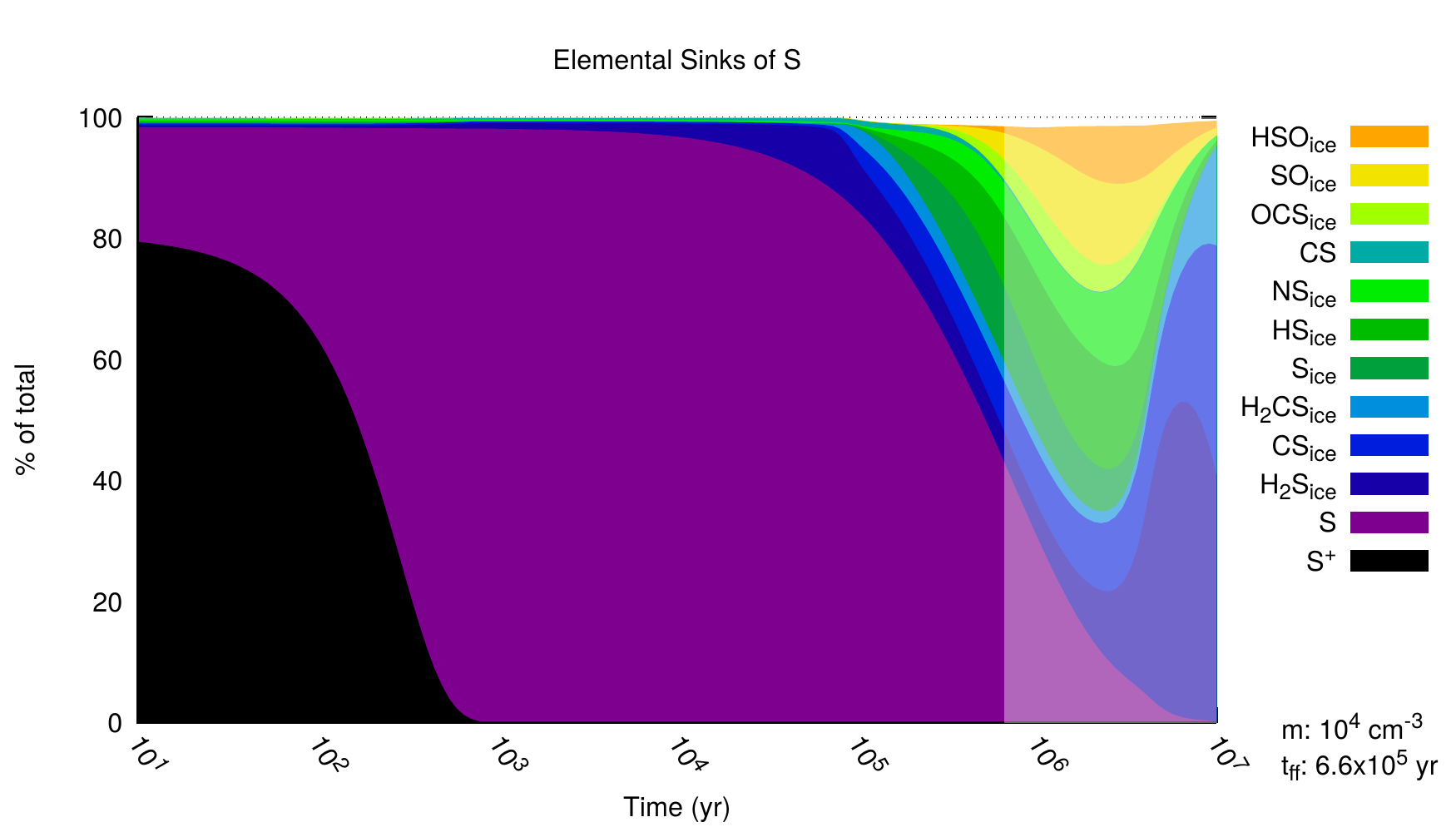}} 
\resizebox{0.82\hsize}{!}{\includegraphics[trim={0 0.5cm 0 1.1cm},clip]{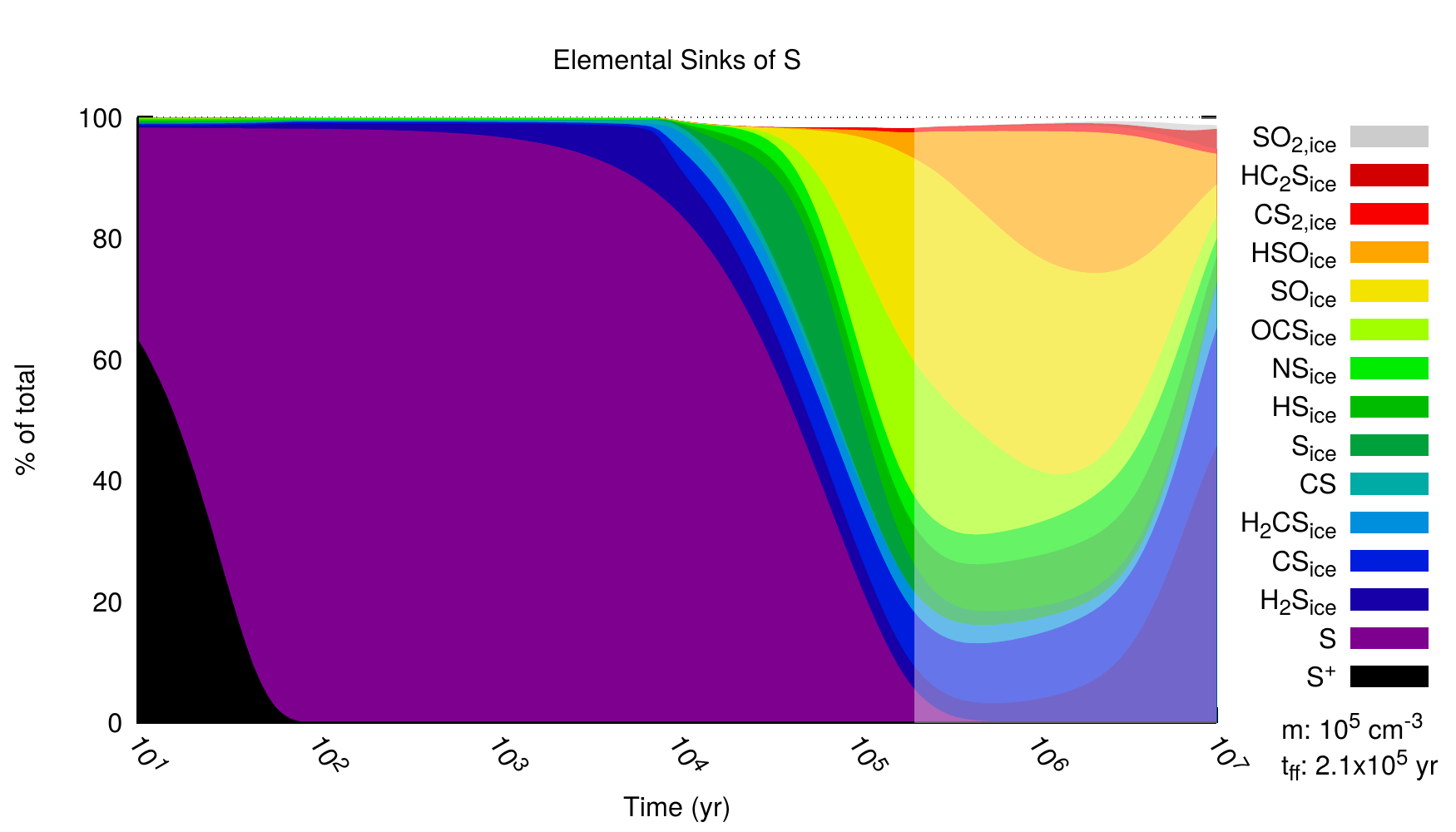}} 
\resizebox{0.82\hsize}{!}{\includegraphics[trim={0 0.1cm 0 1.1cm},clip]{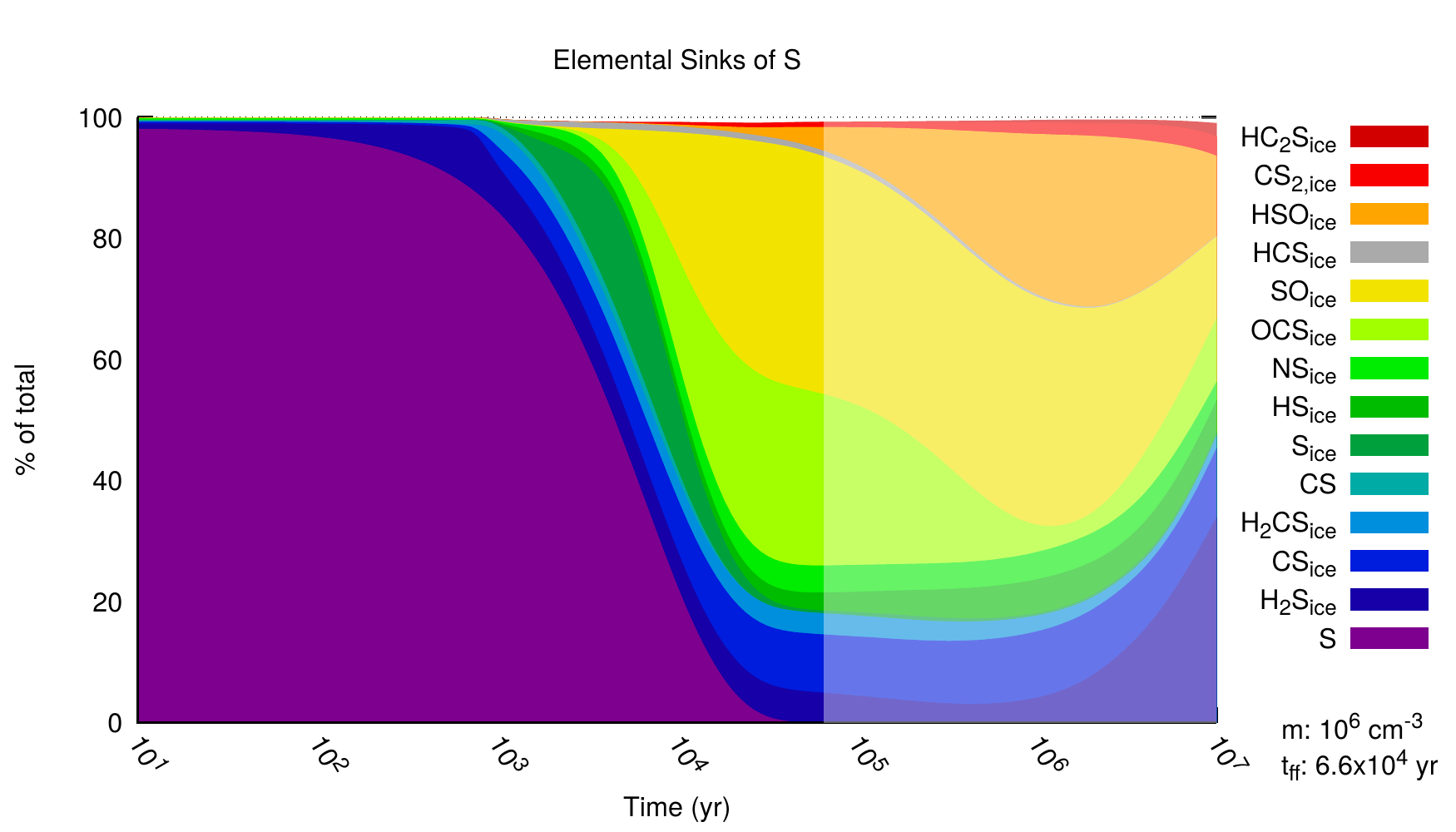}}
\caption{Shown here are the primary S-carrying species from the molecule as a function of time and different volume densities (from $10^4$ cm$^{-3}$, top, to $10^6$ cm$^{-3}$, bottom) during Stage 3. The vertical width of each component represents how much sulfur is accounted for. We have also used partially opaque boxes to highlight the timescale that is defined by the free fall time of a homogeneous cloud to collapse under ideal gravitational conditions ($t_{ff} \approx \sqrt{\nicefrac{3}{2 \pi G \rho_0}}$, i.e. ignoring magnetic fields, angular momentum, and other resistive effects), for reference to other models. The gas number densities, $m$, and approximate free fall time limits, $t_\text{ff}$, are noted for each plot.\label{fig:sulfursinks_dense}}
\end{figure*}

Regarding the early stages of interstellar cloud evolution, we show in Figure \ref{fig:sulfursinks_early} that sulfur remains fully in the atomic ionized form during the diffuse cloud, and it is eventually partially neutralized (ca. 20\%) late into the translucent stage. It is only during the late stage of a more dense environment that it finally naturally depletes ($>99\%$) from the gas-phase, into a diverse spread of simple molecules. To illustrate this, we have collected stacked histogram-like plots in Figure \ref{fig:sulfursinks_dense} showing the molecules that dominate the sulfur budget during Stage 3, where the horizontal axis tracks the time, and the vertical component represents the abundance of each species as a percentage of the total sulfur composition. We have also overlaid these plots with a semi-transparent box to denote the gravitational free-fall time of a cloud at their respective densities. At the left of these plots, it can be seen that sulfur is primarily atomic at early times, but then \ce{H2S} and other grain species become increasingly abundant as freeze-out takes over. At short timescales, S adsorbs onto the grain and is rapidly hydrogenated to \ce{H2S}, however, only to a limited extent of $\sim7\%$. This upper limit agrees very well with the non-detections of solid \ce{H2S} \citep[e.g.][]{Smith1991} toward a number of interstellar environments, and contrast starkly with all other gas/grain astrochemical models. This trend also agrees well with laboratory studies that have shown that solid \ce{H2S} can be rapidly converted to a variety of other species \citep{Garozzo2010, Chen2015}; we discuss some of its chemistry in more detail below (\S\ref{sec:sulfanes}).

An intriguing observation that should also be noted is the detection of solid OCS toward IRC-L1041-2, reported by \citet{Aikawa2012}, yielding a fractional abundance of approximately $1.6\times10^{-6}$ relative to hydrogen. This is an incredibly high fraction of the total abundance of cosmic elemental sulfur, but is indeed a value within the range of our high-density models between gas densities of $10^4$ cm$^{-3}$ and $10^5$ cm$^{-3}$. In this case, there is clear evidence that a significant fraction of the elemental sulfur is trapped in the form of a condensed species, and our model clearly suggests the rest may be found across a variety of other S-bearing organics.

We show in our model that simple freeze-out can account for the appropriate timescale of the gas-phase depletion of sulfur. The major result is that our expanded reaction network produces a significant amount of organo-sulfur molecules, which prevents the build-up of \ce{H2S} on the grain. The direct adsorption of positively-charged ions onto negatively-charged grains has been added. This enhances the freeze-out kinetics by a factor of 50--100\% (not shown here). We speculate that the organo-sulfur chemistry together with the freeze-out can explain why it has been so challenging to reconcile interstellar sulfur depletion from an observational perspective. We predict that sulfur gas-phase depletion does not significantly occur at low densities; however, it does occur rapidly ($\lesssim 10^5$ yrs) at densities higher than $10^4$ cm$^{-3}$, producing solid organo-sulfur molecules that are hard to detect.


\subsection{Comparison to observations}

For the three evolutionary stages of interstellar clouds on which we have based our model, there have been a number of molecular surveys that provide general observational constraints of S-bearing species. These observations have all been summarized below, in Table \ref{table:sulfurdetections}, and are also illustrated in various figures below, alongside related model results. In this section we look at these constraints in more detail through comparisons with our model.

\begin{table}\centering
\caption{Detections and Upper Limits of Sulfur Species in Interstellar Clouds}
\label{table:sulfurdetections}
\begin{tabular}{llll}
\hline
\hline
Species & Frac. Abund. & Env.\tablefootmark{a} & Ref. \\
\hline
\ce{HS}    & $4\times10^{-9}$      & Diffuse   & (1) \\ 
\ce{HS+}   & $\leq4\times10^{-9}$  & Diffuse   & (2) \\ 
\ce{H2S}   & $\sim10^{-9}$         & Diffuse   & (1) \\ 
           & $6\times10^{-9}$      & Transl.   & (3) \\
           & $7\times10^{-10}$     & Dense     & (4) \\
\ce{CS}    & $\leq10^{-9}$         & Diffuse   & (1) \\ 
           & $4\times10^{-9}$      & Transl.   & (6) \\
           & $6.5\times10^{-9}$    & Dense     & (7) \\
\ce{HCS+}  & $2\times10^{-10}$     & Transl.   & (6) \\ 
           & $2.5\times10^{-10}$   & Dense     & (7) \\
\ce{H2CS}  & $\sim3\times10^{-9}$  & Transl.   & (3) \\ 
           & $2.6\times10^{-9}$    & Dense     & (7) \\
\ce{C2S}   & n/a                   & Diffuse   &  \\ 
           & $\leq10^{-9}$         & Transl.   & (9) \\
           & $5\times10^{-9}$      & Dense     & (7) \\
\ce{NS}    & $4\times10^{-10}$     & Dense     & (10) \\ 
\ce{SO}    & $\leq5\times10^{-10}$ & Diffuse   & (1) \\ 
           & $4\times10^{-8}$      & Transl.   & (11) \\
           & $\sim10^{-8}$         & Dense     & (12) \\
\ce{SO+}   & $\sim10^{-9}$         & Transl.   & (3) \\ 
\ce{SO2}   & n/a                   & Diffuse   &  \\ 
           & $<10^{-9}$            & Transl.   & (11) \\
           & $\sim10^{-10}$        & Dense     & (13) \\
\ce{OCS}   & n/a                   & Diffuse   &  \\ 
           & $\sim10^{-9}$         & Transl.   & (3) \\
           & $\sim10^{-9}$         & Dense     & (14) \\
\hline
\end{tabular}
\tablefoot{
\tablefoottext{a}{The environments are classified accordingly: `Diffuse', `Transl.', and `Dense' refer to diffuse ($A_v \leq 1$), translucent ($1 \lesssim A_v \lesssim 2$), and dense ($A_v > 5$) interstellar clouds, respectively.}}
\tablebib{
(1)~\citet{Neufeld2015}; (2)~\citet{Neufeld2012}; (3)~\citet{Turner96sVII}; (4)~\citet{Minh1989}; (5)~\citet{Crocket2014}; (6)~\citet{Turner96sVI}; (7)~\citet{Gratier2016}; (8)~\citet{Minh2011}; (9)~\citet{Turner98sIX}; (10)~\citet{McGonagle1994}; (11)~\citet{Turner95sV}; (12)~\citet{Lique2006}; (13)~\citet{Cernicharo2011}; (14)~\citet{Matthews1987}.}
\end{table}

\subsubsection{Diffuse clouds}

\begin{figure}[!]\centering
\resizebox{\hsize}{!}{\includegraphics[trim={0.23cm 0.15cm 0.35cm 0.6cm},clip]{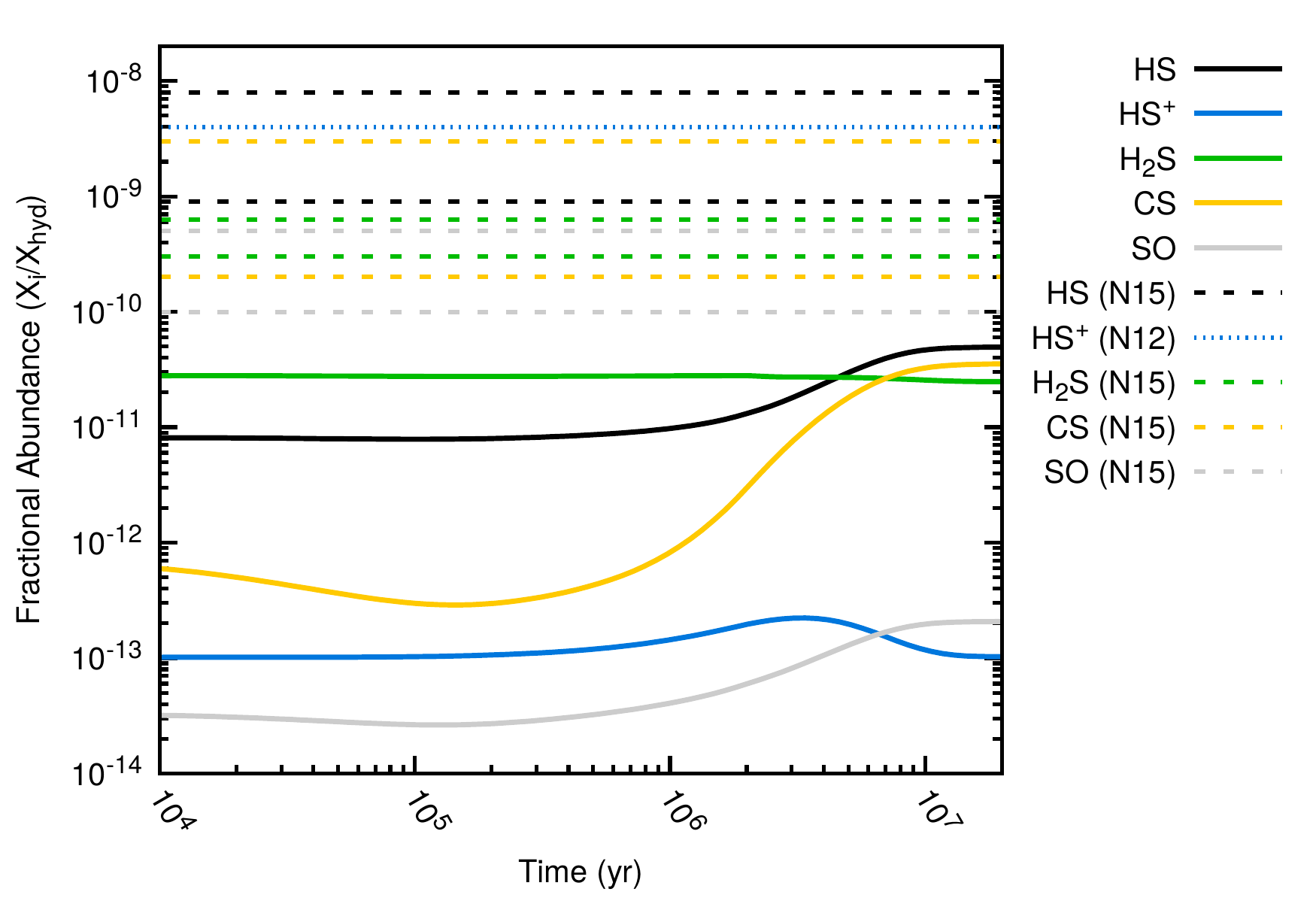}} 
\caption{Comparison of model (solid curves) vs observations (dashes) for a diffuse cloud environment. The thin dashed lines serve as upper limits of observations, and the thick dashed lines function in pairs as a range of observational values. See text for details. The references refer to the following: (N12) \citet{Neufeld2012}; (N15) \citet{Neufeld2015}.\label{fig:stage1_sobs_diffuse}}
\end{figure}

A comparison of the model results against observations of diffuse clouds reported by \citet{Neufeld2012, Neufeld2015} is shown in the plots of Figure \ref{fig:stage1_sobs_diffuse}. Despite a `cosmic' initial abundance of elemental sulfur (i.e. $X_0(S^+) = 1.66\times10^{-5}$), all gas-phase S-bearing species that we focus on are underestimated to varying degrees with respect to observations. From the figure, it can be seen that the observed relative abundances of the species agree in a qualitative manner with the model results. Quantitatively, these species are all under-produced in the model, by a factor of $\gtrsim 10$. The significant underproduction of \ce{HS+}, as well as the underproduction of \ce{HCS+} and \ce{SO+} during the translucent stage (see below) suggests a shortcoming of our simplified physical model for ion chemistry during the early stages of cloud evolution, which we speculate could be due to a number of reasons, such as our simplistic homogeneous medium, our simplistic model of monodisperse dust grains, or our neglect of energy injection via shocks and turbulence dissipation. We do not attempt to improve the model during this early stage, and, instead, refer the reader to previous models of ion chemistry within these environments \citep[e.g.][]{NeufeldDalgarno1989, Cecchi2000, Scappini2000, LePetit2004, Wakelam2008, Neufeld2015, Ivlev2018}.

\subsubsection{Translucent clouds}

\begin{figure}\centering
\resizebox{\hsize}{!}{\includegraphics[trim={0.23cm 0.15cm 0.35cm 0.6cm},clip]{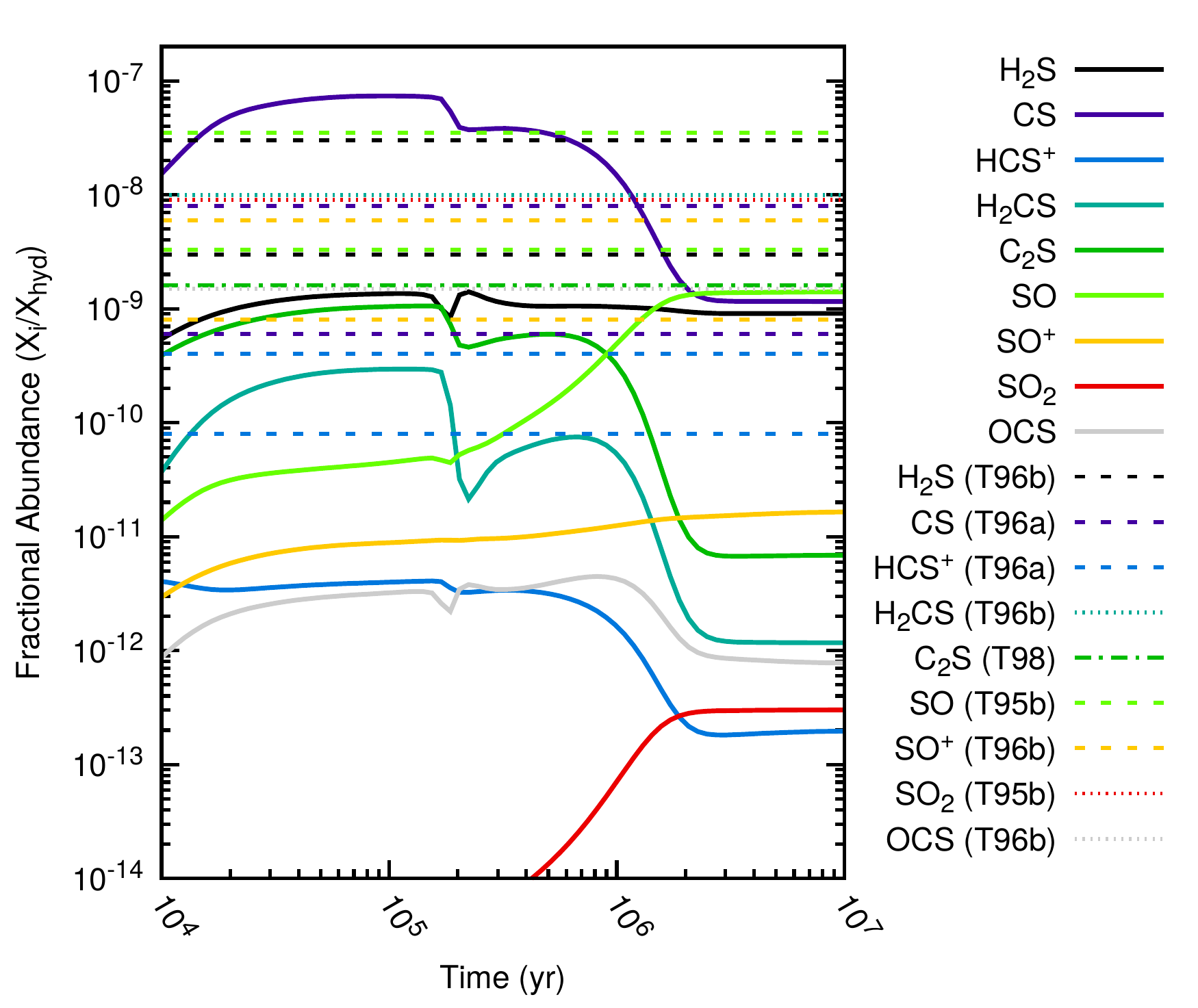}} 
\caption{Comparison of model (solid curves) vs observations (dashed lines) for a translucent cloud environment. See text for details. The references refer to the following: (T95b) \citet{Turner95sV}; (T96a) \citet{Turner96sVI}; (T96b) \citet{Turner96sVII}; (T98) \citet{Turner98sIX}.\label{fig:stage2_sobs_translucent}}
\end{figure}

A thorough survey of many S-bearing species in translucent cloud environments has been presented in \citet{Turner95sV, Turner96sVI, Turner96sVII} and \citet{Turner98sIX}. These reports have provided self-consistent constraints that we can use for comparison with our model, which we show in Figure \ref{fig:stage2_sobs_translucent}. In general, the model performs much better at predicting gas-phase abundances of S-bearing species for the translucent stage compared to the diffuse stage, however, most species remain underpredicted. Of these species, \ce{CS} is predicted at the highest abundance within the model, at a level of nearly $10x$ above the upper range of abundances reported by \citet{Turner96sVII} before $t = 1$ Myr, but then settling within this range afterwards. The two species observed at the highest abundances by \citet{Turner95sV, Turner96sVI, Turner96sVII} and \citet{Turner98sIX} are \ce{H2S} and \ce{SO}, which are both underpredicted by only a factor of a few around $t = 1$ Myr compared to observations. \ce{C2S} is predicted at nearly the value reported by \citet{Turner98sIX}, however it was detected in only a small number of sources with respect to total number, and this value therefore serves as more like an upper limit. The rest of the species---\ce{H2CS}, \ce{SO+}, \ce{OCS}, \ce{HCS+} and \ce{SO2}---are all severely underpredicted by the model compared to their ranges of reported abundances. Again, we speculate that our underpredictions might simply be due to our overly-simplistic model of low-density regions that is lacking certain processes important for diatomics and molecular ions.

\subsubsection{Dense environments}

\begin{figure}[!]\centering
\resizebox{\hsize}{!}{\includegraphics[trim={0.23cm 0.15cm 0.35cm 0.6cm},clip]{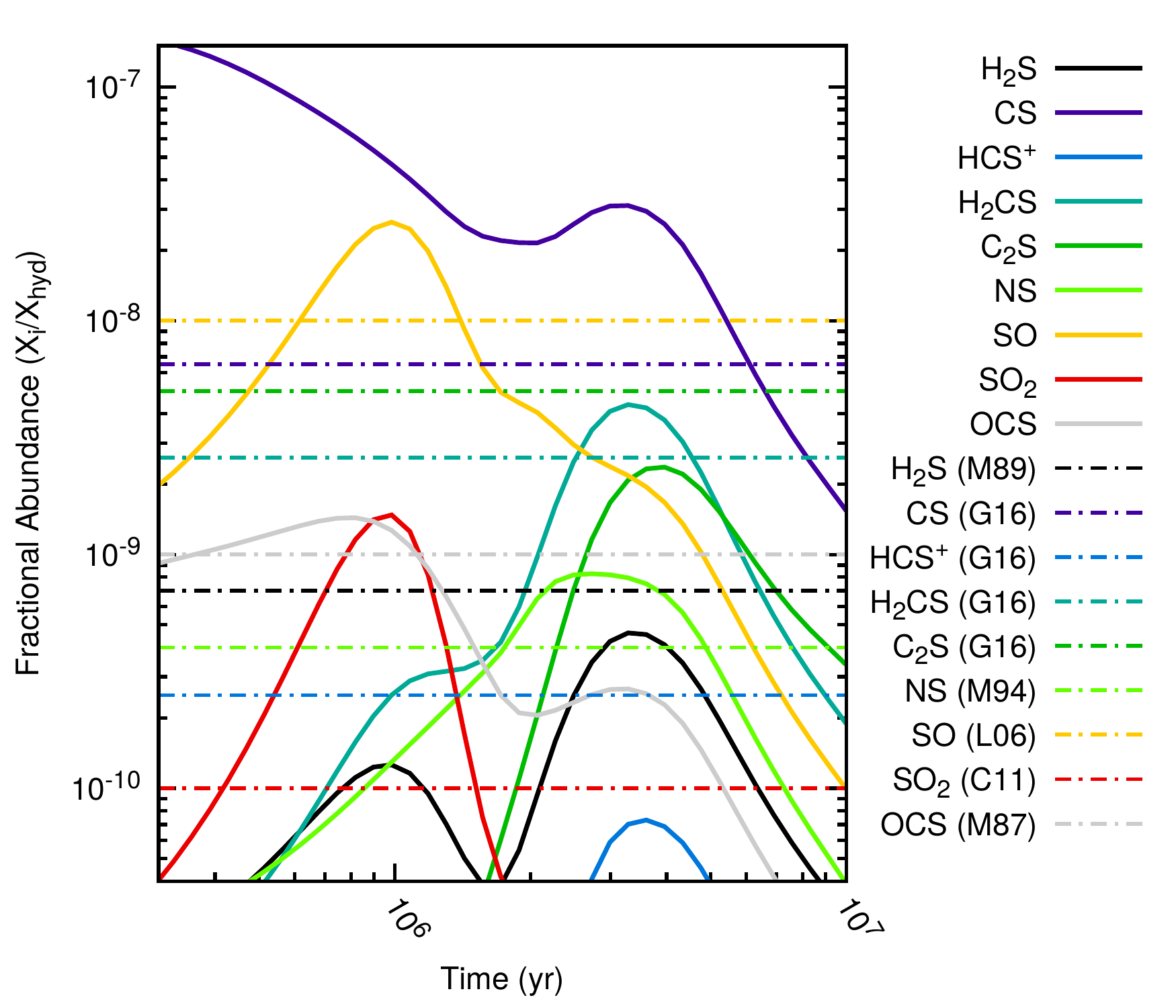}} 
\caption{Comparison of model (solid curves) vs observations (dashed lines) for a dense cloud environment. The observed values all pertain to the cold, dark cloud TMC-1 ($\sim10^4$ cm$^{-3}$), from the following references: (C11) \citet{Cernicharo2011}; (G16) \citet{Gratier2016}; (L06) \citet{Leurini2006}; (M87) \citet{Matthews1987}; (M89) \citet{Minh1989}; (M94) \citet{McGonagle1994}. See text for more details.\label{fig:stage3_sobs_dense}}
\end{figure}

For our comparisons to dense clouds, it was desired to focus on a single environment, though few systematic studies of large sets of S-bearing species could be identified. It was necessary, therefore, to combine a number of datasets. TMC-1 is a cold, dark cloud that has been very well-characterized regarding both its chemical and physical characteristics, making it one of the best matches for our Stage 3 model at a density of $10^4$ cm$^{-3}$. These observational constraints are shown alongside our model results through the time plot in Figure \ref{fig:stage3_sobs_dense}.

For the model trial at a density of $10^4$ cm$^{-3}$, all molecules are predicted in very good agreement with observations at times greater than $10^6$ years. Some species are clearly very sensitive to the ``chemical age'' of the model, particularly \ce{SO2}, which has an enhanced abundance at early times of the dense stage that better agrees with translucent cloud observations \citep{Turner95sV} rather than towards TMC-1 \citep{Cernicharo2011}. On the other hand, the relative abundances of \ce{SO} and \ce{SO2} (\nicefrac{\ce{SO}}{\ce{SO2}} $\approx$ 5--10) are consistent with the range that is typical for many other interstellar observations \citep[][e.g.]{vanderTak2003}, though their observed abundances do appear to vary significantly from source to source.

We can also compare the relative abundance ratios of various S-bearing species against more commonly observed molecular species. In Table \ref{table:sulfurratios_gas} we compare the gas-phase abundances of several S-bearing species as a ratio with respect to \ce{CO}, where the values have been chosen at times that give approximately the peak abundance for the species shown in Figure \ref{fig:stage3_sobs_dense}. These abundance ratios vary significantly across three interstellar environments, but each modeled density of our dense stage gives reasonable agreement with these values. Furthermore, \ce{OCS} has been a well-studied component of interstellar ices via its absorption band near 2040 cm$^{-1}$. It is the only S-bearing species detected in interstellar ices, and our model predicts that its formation is driven primarily by S-atom addition to CO (see Appendix \ref{sec:ocs} for more details). In Table \ref{table:sulfurratios_OCSice} we compare a summary of \ce{OCS_{ice}} observations with respect to solid CO toward a variety of protostars, where our selected points in time match those of the previous table (Table \ref{table:sulfurratios_gas}).

\begin{table}\centering
\caption{Abundance Ratios of Select Gas-Phase Species, \ce{X_{gas}} vs \ce{CO_{gas}}}
\label{table:sulfurratios_gas}
\begin{tabular}{llllll}
\hline
\hline
Environment & \ce{H2S} & \ce{OCS} & \ce{H2CS} & \ce{SO} & \ce{SO2} \\
\hline
TMC-1\tablefootmark{a} & 0.0009 & 0.001 & 0.003   & 0.01  & 0.0001 \\
L134N\tablefootmark{b} & 0.001  & 0.002 & 0.00008 & 0.025 & 0.0005 \\
Orion\tablefootmark{b} & 0.1    & 0.05  & 0.001   & 0.05  & 0.06 \\
\hline
$n=10^4$ cm$^{-3}$  & 0.0008 & 0.0004 & 0.003 & 0.02 & 0.0002 \\
$t=2$ Myr     & & & & & \\
$n=10^5$ cm$^{-3}$  & 0.01   & 0.003  & 0.02  & 0.08 & 0.003 \\
$t=0.3$ Myr   & & & & & \\
$n=10^6$ cm$^{-3}$  & 0.05   & 0.03   & 0.01  & 0.7  & 0.06 \\
$t=0.03$ Myr  & & & & & \\
\hline
\end{tabular}
\tablefoot{Values are in terms of percentage with respect to \ce{CO_{gas}}. Observed values are determined from the following:
\tablefoottext{a}{Observed values for the S-bearing species are based on those referenced in Table \ref{table:sulfurdetections} and Figure \ref{fig:stage3_sobs_dense}, and scaled with respect to a value of $f(CO) = 8 \times 10^{-5}$ as per \citet{Irvine1987}.}
\tablefoottext{b}{All values are taken from Table 1 of \citet{Charnley2001}.}
}
\end{table}

\begin{table}\centering
\caption{Abundance Ratios of Interstellar \ce{OCS_{ice}} vs \ce{CO_{ice}}}
\label{table:sulfurratios_OCSice}
\begin{tabular}{ll}
\hline
\hline
Environment & Abundance (\%) \\
\hline
W33A & 5 \\
AFGL 989 & 0.8 \\
Mon R2 IRS2 & 0.65 \\
AFGL 961E & $<1$ \\
AFGL 490 & $<2.6$ \\
NGC 2024 IRS 2 & $<0.8$ \\
OMC 2 IRS 3 & $<1.6$ \\
Elias 16 & $<0.8$ \\
\hline
$n=10^4$ cm$^{-3}$  & 0.7 \\
$t=2$ Myr      & \\
$n=10^5$ cm$^{-3}$  & 4 \\
$t=0.3$ Myr    & \\
$n=10^6$ cm$^{-3}$  & 5 \\
$t=0.03$ Myr   & \\
\hline
\end{tabular}
\tablefoot{Observed values are in terms of percentage with respect to \ce{CO_{ice}}, and are taken from Table 2 of \citet{Palumbo1997}.}
\end{table}


\section{Discussion} \label{sec:discussion}

Our gas/grain model incorporates a number of general updates to the parent OSU gas/grain model \citep{Garrod2008} pertaining to photochemistry so that the model may also be used at earlier stages of cloud evolution, namely low-$A_\mathrm{v}$ conditions ($0.5 \lesssim A_\mathrm{v}~(mag) \lesssim 2$). These photochemistry updates have allowed us to include an explicit third stage for the aging of a diffuse cloud, and the inclusion of the earlier stages and cosmic abundances provides the opportunity to check for elements of sulfur depletion or unexpected effects caused by our changes. In general terms, however, our model predictions do not perform well for minor gas-phase species at earlier stages, but we are confident that the underprediction of many gas-phase sulfur-containing species does not significantly affect the major results regarding sulfur depletion during the dense stage. This has been confirmed by alternatively running the later stages with initial abundances of select species fixed to better match observational values of the preceding stages, but their abundances rapidly (within $10^2$ yrs) converge to the same results at intermediate- and late-times (not shown here).

We have also adopted a higher binding energy for \ce{O_{surf}} and a lower (compared to the previous OSU model) binding energy for \ce{NH3{}_{surf}} under the guidance of recent laboratory studies, and these differences will affect basic ice chemistry of water and ammonia at later stages.

Our model also considers a number of novel approaches with regard to sulfur chemistry, to better bring it into alignment with other chemistries, such as carbon, nitrogen, and oxygen. In general terms, our new model provides the most expansive astrochemical network for sulfur to-date. Unlike many previous models, we have made broad updates across the entire sulfur reaction network, and have sourced as much as possible all previous laboratory and theoretical studies of sulfur chemistry in an unbiased manner, including the correction of many erroneous rate coefficients and activation barriers that were used in the past prior to these more recent studies.

Regarding sulfur allotropes (sulfur chains/rings), we have made significant updates to the network and molecular parameters, to be consistent with the literature about solid sulfur, so that they are treated just like any other chemical species and may interact with the reaction network through a number of processes. This is unlike other studies which have used a more heuristic approach with only a subset of reactions/processes to test for their importance under specific conditions \citep[e.g.][]{Wakelam2005, Woods2015}. On the other hand, these heavy species do not appear active in our model, at such a low temperature and quiescent physics.

We have also newly added a number of species related to \ce{CS2} and the \ce{C_xS} carbon-sulfur chains, many of which have been both studied in the laboratory and detected in interstellar environments, but not yet considered in a general astrochemical model. In the case of \ce{CS2}, it appears to be much more important in cometary ices than interstellar environments, though its interstellar significance may be underestimated due to its detection being significantly impeded without a permanent dipole moment. Nevertheless, we have included it and many of its hydrogenated derivatives. As is the case with the oxygen counterparts, the singly-hydrogenated forms of \ce{C2S} and \ce{C3S} are expected to be important molecules relative to their parents. With the exception of ethyl mercaptan (\ce{CH3CH2SH}), our model therefore includes all known interstellar sulfur-bearing species to-date, and is therefore the most complete astrochemical reaction network to-date.

Finally, unlike recent models of interstellar sulfur, we show that sulfur can be depleted from its gas-phase cosmic abundance onto the grain at late-times during the dense molecular phase, and this grain composition is not dominated by \ce{H2S}. To prevent this unrealistically large build-up of solid \ce{H2S}, it was important to include its enhanced photodissociation on the grain (compared to its gas-phase rate), its photodesorption under conditions of low-A$_\text{V}$, and the extensive grain reactions that provide more routes to a generally more diverse range of sulfur chemistry.


\section{Conclusion} \label{sec:conclusions}

We have modeled sulfur chemistry using an expanded gas/grain astrochemical model across the evolution of an interstellar cloud. Our model suggests that the vast majority of the sulfur eventually becomes locked up in the condensed phase and across a variety of simple organic molecules, which may explain the ongoing challenge of detecting enough gas-phase sulfur-bearing species to account for its cosmic abundance. Our model demonstrates that an enhanced gas/grain astrochemical model of sulfur chemistry can provide insight into the observed sulfur depletion from the cosmic gas-phase abundance, without breaking agreement with many gas-phase S-bearing species which models have generally already succeeded.

Our model suggests that sulfur depletion from the gas phase does occur under dark cloud conditions if the cloud has an age greater than $10^6$ years. This process is caused by rapid freeze-out at enhanced densities, which also results in highly non-equilibrium chemistry that appears to be highly sensitive to the gas density and timescale of the dense cloud stage, resulting in a variable molecular inventory that may present a challenge in modeling specific interstellar environments.

Our model provides no evidence of sulfur chains/rings forming a significant refractory component, and alternatively suggests that elemental sulfur is converted to a range of simple, stable organo-sulfur molecules that are primarily formed there. Many of these species resemble those which have been detected in cometary environments, however some differences also exist, such as our negligible reservoir of solid \ce{H2S}. During the preparation of this manuscript, a number of new reports related to interstellar sulfur molecules have showed up in the literature (i.e. \ce{HCS}, \ce{S2H}, \ce{HSO}, and \ce{NS+}), and their detections/upper limits have also been found to agree very well with our model. This suggests that our work is a significant step forward for astrochemical models of sulfur, and we hope we have provided enough information to serve as a stepping stone for future studies.

We have provided a summary of key formation reactions behind the full range of S-bearing interstellar molecules, many of which are new to a gas/grain astrochemical model. The variety of reactions presented here reflects the diverse chemistry of sulfur, highlighting the ability of sulfur to react with a variety of types of molecules/elements---including itself---and across a range of oxidation states. Not surprisingly, many reaction rates presented here are uncertain, leaving room for much improvement by way of laboratory and/or theoretical studies.


\begin{acknowledgements}
The authors would like to thank Profs. Guillermo Mu\~{n}oz Caro and Maria Palumbo for conversations about laboratory studies of sulfur-containing ices, Prof. Susanna Widicus Weaver about gas/grain modeling, and Profs. John Black and David Neufeld about observations. JL would also like to thank Dr. Wing-Fai Thi for helpful feedback during the development of the astrochemical code.
\end{acknowledgements}


\begin{appendix}

\section{Updated binding energies} \label{app:be}
In Table \ref{table:binde}, we show all the adopted binding energies for all species that may interact/reside on the grain. These values have been the subject of many recent studies, and we have tried to implement a number of suggestions from the literature. In the many cases where the values have only been guessed, we have included a note/formula behind the estimation. In most cases, it has been our experience that similar formulae for many complex molecules are consistent with measured/theoretical values.

{\longtab[1]{
\begin{longtable}{llll}
\caption{Binding energies and heats of formation for all grain species. \label{table:binde}} \\
\hline
\hline
Species & Binding Energy, & Heat of Formation, & Remark\tablefootmark{a} \\
 & $E_{bind}$ (K) & $\Delta H_f$ (\nicefrac{kcal}{mol}) &  \\
\hline
\endfirsthead
\caption{Binding energies and heats of formation for all grain species. (cont'd)} \\
\hline
Species & $E_{bind}$ (K) & $\Delta H_f$ (\nicefrac{kcal}{mol}) & Remark\tablefootmark{a} \\
\hline
\endhead
\hline
\endfoot
\hline
\endlastfoot
\ce{H}               &  450     &    +51.63   &  \\
\ce{H2}              &  430     &       0.0   &  \\
\ce{He}              &  100     &       0.0   &  \\
\ce{C}               &  800     &   +169.98   & $\Delta H_f$ from (1) \\
\ce{N}               &  800     &   +112.53   &  \\
\ce{O}               &  1764    &    +58.99   & $E_{bind}$ from (2) \\
\ce{Na}              &  11800   &    +25.69   &  \\
\ce{Mg}              &  5300    &    +34.87   &  \\
\ce{Si}              &  2700    &    +106.6   &  \\
\ce{P}               &  1100    &    +75.62   &  \\
\ce{S}               &  1100    &    +65.66   &  \\
\ce{Cl}              &  3000    &    +28.99   & $E_{bind}$ from (3) \\
\ce{Fe}              &  4200    &    +99.30   &  \\
\ce{CH}              &  925     &   +142.00   & C + \nicefrac{1}{4}*(\ce{CH4}-C) \\
\ce{CH2}             &  1050    &    +93.35   & C + \nicefrac{2}{4}*(\ce{CH4}-C) \\
\ce{CH3}             &  1175    &    +34.82   & C + \nicefrac{3}{4}*(\ce{CH4}-C) \\
\ce{CH4}             &  1300    &    -17.89   &  \\
\ce{NH}              &  1533    &    +85.92   & N + \nicefrac{1}{3}*(\ce{NH3}-N) \\
\ce{NH2}             &  2267    &    +45.50   & N + \nicefrac{2}{3}*(\ce{NH3}-N) \\
\ce{NH3}             &  3000    &      -9.3   & $E_{bind}$ from (4), $\Delta H_f$ from (1) \\
\ce{OH}              &  2850    &     +8.85   & \nicefrac{1}{2}*\ce{H2O} \\
\ce{H2O}             &  5700    &    -57.10   &  \\
\ce{O2}              &  930     &       0.0   & $E_{bind}$ from (5) \\
\ce{O3}              &  1833    &    +34.10   & $E_{bind}$ from (5) \\
\ce{HO2}             &  4614    &     +0.50   & O + \ce{OH} \\
\ce{H2O2}            &  5700    &    -32.53   & 2*\ce{OH} \\
\ce{NaH}             &  12250   &    +29.70   & Na + H \\
\ce{NaOH}            &  14650   &    -47.27   & Na + \ce{OH} \\
\ce{NaCN}            &  13400   &     +22.5   & Na + \ce{CN}, $\Delta H_f$ from (6) \\
\ce{FeH}             &  4650    &    +117.2   & Fe + H, $\Delta H_f$ from (7) \\
\ce{FeS}             &  5300    &     +83.8   & Fe + S, $\Delta H_f$ from (7) \\
\ce{MgH}             &  5750    &    +40.40   & Mg + H \\
\ce{MgH2}            &  6200    &     -18.2   & Mg + 2*H, $\Delta H_f$ from (8) \\
\ce{SiH}             &  3150    &    +90.02   & Si + H \\
\ce{SiH2}            &  3600    &    +62.75   & Si + 2*H \\
\ce{SiH3}            &  4050    &    +35.47   & Si + 3*H \\
\ce{SiH4}            &  4500    &     +8.20   & Si + 4*H \\
\ce{SiC}             &  3500    &    +177.1   & Si + C, $\Delta H_f$ from (6) \\
\ce{SiCH}            &  3625    &             & Si + CH \\
\ce{SiCH2}           &  3750    &             & Si + \ce{CH2} \\
\ce{SiCH3}           &  3875    &             & Si + \ce{CH3} \\
\ce{SiC2}            &  4300    &    +152.0   & \ce{SiC} + C, $\Delta H_f$ from (6) \\
\ce{SiC2H}           &  4837    &             & Si + \ce{C2H} \\
\ce{SiC2H2}          &  5287    &             & Si + \ce{HC2H} \\
\ce{SiC3}            &  5100    &    +228.8   & Si + \ce{C3}, $\Delta H_f$ from (6) \\
\ce{SiC3H}           &  5637    &             & Si + \ce{C3H} \\
\ce{SiC4}            &  5900    &    +216.9   & \ce{SiC} + \ce{C3}, $\Delta H_f$ from (6) \\
\ce{SiN}             &  3500    &    +89.00   & Si + N \\
\ce{HNSi}            &  4233    &             & Si + \ce{NH} \\
\ce{SiNC}            &  4300    &             & Si + \ce{CN} \\
\ce{SiO}             &  4464    &    -24.00   & Si + O \\
\ce{SiO2}            &  6228    &    -73.00   & 2*\ce{SiO} - Si \\
\ce{H2SiO}           &  5364    &    -999.9   & 2*\ce{SiH} + \ce{SiO} - 2*Si \\
\ce{SiS}             &  3800    &     +25.3   & Si + S, $\Delta H_f$ from (9) \\
\ce{PH}              &  1550    &    +60.60   & P + H \\
\ce{PH2}             &  2000    &    +30.10   & P + 2*H \\
\ce{PN}              &  1900    &    +25.04   & P + N \\
\ce{PO}              &  2864    &     -5.63   & P + O \\
\ce{HCP}             &  2025    &    +35.83   & \ce{CH} + \ce{CP} - C \\
\ce{HPO}             &  3314    &    -12.70   & P + O + H \\
\ce{S2}              &  550     &     +50.2   & 2*S, $\Delta H_f$ from (10) \\
\ce{S3}              &  1100    &     +49.7   & \ce{S2} + \nicefrac{1}{2}*S, $\Delta H_f$ from (10) \\
\ce{S4}              &  1650    &     +47.8   & \ce{S2} + \nicefrac{2}{2}*S, $\Delta H_f$ from (10) \\
\ce{S5}              &  2200    &     +36.0   & \ce{S2} + \nicefrac{3}{2}*S, $\Delta H_f$ from (10) \\
\ce{S6}              &  2750    &     +31.2   & \ce{S2} + \nicefrac{4}{2}*S, $\Delta H_f$ from (10) \\
\ce{S7}              &  3300    &     +31.0   & \ce{S2} + \nicefrac{5}{2}*S, $\Delta H_f$ from (10) \\
\ce{S8}              &  3850    &     +25.1   & \ce{S2} + \nicefrac{6}{2}*S, $\Delta H_f$ from (10) \\
\ce{HS}              &  1372    &     +33.9   & \nicefrac{1}{2}*\ce{H2S} \\
\ce{H2S}             &  2743    &     -4.23   &  \\
\ce{HSO}             &  3314    &       0.0   & \ce{SO} + H, $\Delta H_f$ from (11) \\
\ce{S2H}             &  1000    &     +22.0   & \ce{S2} + H, $\Delta H_f$ from (12) \\
\ce{H2S2}            &  3193    &      +3.8   & \ce{H2S} + H \\
\ce{S2O}             &  2314    &     -12.9   & \ce{S2} + O, $\Delta H_f$ from (13) \\
\ce{CH2SH}           &  2096    &     +37.7   & \ce{CH3SH} - H, $\Delta H_f$ from (14) \\
\ce{CH3S}            &  3250    &     +29.7   & \ce{H2CS} + H, $\Delta H_f$ from (15) \\
\ce{CH3SH}           &  2546    &     -5.46   & \ce{CH3} + \ce{HS}, $\Delta H_f$ from (16) \\
\ce{CS3}             &  1900    &             & C + \ce{S3} \\
\ce{ClH}             &  3450    &    -22.06   & Cl + H \\
\ce{ClC}             &  3800    &   +120.00   & Cl + C \\
\ce{ClO}             &  4764    &    +24.19   & Cl + O \\
\ce{CP}              &  1900    &   +107.53   & C + P \\
\ce{CCP}             &  2700    &    +153.1   & \ce{C2} + P, $\Delta H_f$ from (17) \\
\ce{C3P}             &  3500    &    +174.7   & \ce{C2P} + C, $\Delta H_f$ from (17) \\
\ce{C4P}             &  4300    &    +199.1   & \ce{C3P} + C, $\Delta H_f$ from (17) \\
\ce{CH2PH}           &  2600    &    -67.40   & \ce{CH2} + \ce{PH} \\
\ce{HCCP}            &  3150    &             & \ce{CCP} + H \\
\ce{CO}              &  1150    &    -27.20   &  \\
\ce{CO2}             &  2575    &    -93.97   &  \\
\ce{HCO}             &  1600    &    +10.40   & \ce{CO} + H \\
\ce{HOC}             &  3650    &    +10.40   & C + \ce{OH} \\
\ce{H2CO}            &  2050    &    -27.70   & \ce{HCO} + H \\
\ce{HCOH}            &  4634    &    -27.70   & \ce{CH3OH} - 2*H \\
\ce{CH2OH}           &  5084    &     -4.10   & \ce{CH3OH} - H, $\Delta H_f$ from (9) \\
\ce{CH3O}            &  2500    &     +4.10   & \ce{H2CO} + H \\
\ce{CH3OH}           &  5534    &    -48.00   &  \\
\ce{N2}              &  1000    &       0.0   &  \\
\ce{N2H}             &  1450    &    +76.20   & \ce{N2} + H \\
\ce{N2H2}            &  3067    &    +50.90   & \ce{NH2} + N \\
\ce{NO}              &  2564    &    +21.58   & N + O \\
\ce{HNO}             &  3014    &    +23.80   & H + \ce{NO} \\
\ce{HON}             &  3650    &    +23.80   & N + \ce{OH} \\
\ce{NO2}             &  4328    &     +7.91   & N + 2*O \\
\ce{N2O}             &  3364    &    +19.61   & 2*N + O \\
\ce{NS}              &  1900    &    +63.00   & N + S \\
\ce{NH2CN}           &  3867    &    +33.67   & \ce{NH2} + \ce{CN}, $\Delta H_f$ from (18) \\
\ce{SO}              &  2864    &     +1.12   & S + O \\
\ce{SO2}             &  3405    &    -70.33   &  \\
\ce{SO3}             &  5169    &     -93.3   & \ce{SO2} + O, $\Delta H_f$ from (13) \\
\ce{CN}              &  1600    &    +106.4   & C + N, $\Delta H_f$ from (17) \\
\ce{HCN}             &  2050    &    +32.30   & \ce{CN} + H \\
\ce{HNC}             &  2050    &     +46.5   & \ce{HCN}, $\Delta H_f$ from (6) \\
\ce{CNH2}            &  2500    &    +55.70   & \ce{CN} + 2*H \\
\ce{H2CN}            &  2500    &    +55.70   & \ce{CN} + 2*H \\
\ce{CS}              &  1900    &     +72.4   & C + S, $\Delta H_f$ from (17) \\
\ce{CS2}             &  3800    &     +27.7   & 2*\ce{CS} - C, $\Delta H_f$ from (13) \\
\ce{CSSH}            &  4250    &             & \ce{CS2} + H \\
\ce{HCSSH}           &  3722    &             & \ce{HCS} + \ce{HS} \\
\ce{HCOSH}           &  2972    &     -30.0   & \ce{HCO} + \ce{HS}, $\Delta H_f$ from (19) \\
\ce{HCS}             &  2350    &    +71.70   & \ce{CS} + H \\
\ce{H2CS}            &  2800    &    +24.30   & \ce{HCS} + H, $\Delta H_f$ from (19) \\
\ce{C2S}             &  1075    &   +141.31   & C + \nicefrac{1}{2}*S2, $\Delta H_f$ from (17) \\
\ce{C3S}             &  1875    &   +136.29   & \ce{C2S} + C, $\Delta H_f$ from (17) \\
\ce{C4S}             &  2675    &   +196.96   & \ce{C3S} + C, $\Delta H_f$ from (17) \\
\ce{C5S}             &  3475    &   +190.68   & \ce{C4S} + C, $\Delta H_f$ from (17) \\
\ce{HC2S}            &  1525    &             & \ce{C2S} + H \\
\ce{HC3S}            &  2325    &             & \ce{C3S} + H \\
\ce{OCN}             &  3364    &     +30.6   & O + C + N, $\Delta H_f$ from (1) \\
\ce{HNCO}            &  3814    &    -33.36   & \ce{OCN} + H, $\Delta H_f$ from (6) \\
\ce{HOCN}            &  3814    &      -4.4   & \ce{OCN} + H, $\Delta H_f$ from (6) \\
\ce{OCS}             &  2888    &    -33.90   &  \\
\ce{OCS2}            &  3988    &             & \ce{OCS} + S \\
\ce{HSCN}            &  2972    &     +38.3   & \ce{HS} + \ce{CN}, $\Delta H_f$ from (6) \\
\ce{HNCS}            &  3433    &     +25.0   & \ce{NH} + \ce{CS}, $\Delta H_f$ from (6) \\
\ce{HCNS}            &  2825    &     +59.4   & \ce{CH} + \ce{NS}, $\Delta H_f$ from (6) \\
\ce{NH2CS}           &  4167    &             & \ce{NH2} + \ce{CS} \\
\ce{NH2CHS}          &  4617    &             & \ce{NH2} + \ce{HCS} \\
\ce{NH2CH2SH}        &  4688    &             & \ce{NH2} + \ce{CH2} + \ce{HS} \\
\ce{C2}              &  1600    &   +194.43   & 2*C, $\Delta H_f$ from (17) \\
\ce{C3}              &  2400    &   +244.79   & 3*C, $\Delta H_f$ from (17) \\
\ce{C4}              &  3200    &   +261.33   & 4*C, $\Delta H_f$ from (17) \\
\ce{C5}              &  4000    &   +305.82   & 5*C, $\Delta H_f$ from (17) \\
\ce{C6}              &  4800    &   +310.38   & 6*C, $\Delta H_f$ from (17) \\
\ce{C7}              &  5600    &   +356.50   & 7*C, $\Delta H_f$ from (17) \\
\ce{C8}              &  6400    &   +363.74   & 8*C, $\Delta H_f$ from (17) \\
\ce{C9}              &  7200    &   +373.33   & 9*C \\
\ce{C10}             &  8000    &   +420.89   & 10*C \\
\ce{C2H}             &  2137    &    +148.0   & \ce{C2H2} - H, $\Delta H_f$ from (17) \\
\ce{HC2H}            &  2587    &    +54.19   &  \\
\ce{H2C2H}           &  3037    &    +71.00   & \ce{C2H2} + H \\
\ce{H2C2H2}          &  3487    &    +12.54   & \ce{C2H2} + 2*H \\
\ce{CH3CH2}          &  3937    &    +28.40   & \ce{C2H2} + 3*H \\
\ce{CH3CH2CHO}       &  3825    &    -45.00   & \ce{CH3CHO} + \ce{CH2} \\
\ce{CH3CH3}          &  4387    &    -20.04   & \ce{C2H5} + H \\
\ce{C3H}             &  2937    &    +173.6   & \ce{C2H} + C, $\Delta H_f$ from (17) \\
\ce{C3H2}            &  3387    &   +152.59   & \ce{C3H} + H, $\Delta H_f$ from (17) \\
\ce{C3H3}            &  3837    &    +81.00   & \ce{C3H} + 2*H \\
\ce{C3H4}            &  4287    &    +44.32   & \ce{C3H} + 3*H \\
\ce{C3H5}            &  4737    &    +40.90   & \ce{C3H} + 4*H \\
\ce{C3H6}            &  5187    &     +4.88   & \ce{C3H} + 5*H \\
\ce{C3H7}            &  5637    &    +22.00   & \ce{C3H} + 6*H \\
\ce{C3H8}            &  6087    &    -25.02   & \ce{C3H} + 7*H \\
\ce{C4H}             &  3737    &    +194.8   & \ce{C3H} + C, $\Delta H_f$ from (17) \\
\ce{HC4H}            &  4187    &   +110.69   & \ce{C4H} + H \\
\ce{C4H3}            &  4637    &   +110.66   & \ce{C4H} + 2*H \\
\ce{C4H4}            &  5087    &    +70.40   & \ce{C4H} + 3*H \\
\ce{C4H5}            &  5537    &    +73.00   & \ce{C4H} + 4*H \\
\ce{C4H6}            &  5987    &    +39.48   & \ce{C4H} + 5*H \\
\ce{C5H}             &  4537    &    +224.7   & \ce{C4H} + C \\
\ce{C5H2}            &  4987    &   +185.28   & \ce{C5H} + H, $\Delta H_f$ from (17) \\
\ce{C5H3}            &  5437    &   +130.52   & \ce{C5H} + 2*H \\
\ce{C5H4}            &  5887    &   +115.00   & \ce{C5H} + 3*H \\
\ce{CH3C4H}          &  5887    &   +101.00   & \ce{C5H4} \\
\ce{C6H}             &  5337    &    +247.6   & \ce{C5H} + C, $\Delta H_f$ from (17) \\
\ce{C6H2}            &  5787    &   +161.68   & \ce{C6H} + H, $\Delta H_f$ from (17) \\
\ce{C6H3}            &  6237    &   +161.38   & \ce{C6H} + 2*H \\
\ce{C6H4}            &  6687    &   +128.60   & \ce{C6H} + 3*H \\
\ce{C6H6}            &  7587    &    +19.82   & \ce{C6H} + 5*H \\
\ce{C7H}             &  6137    &    +279.7   & \ce{C6H} + C, $\Delta H_f$ from (17) \\
\ce{C7H2}            &  6587    &   +226.69   & \ce{C7H} + H \\
\ce{C7H3}            &  7037    &   +182.50   & \ce{C7H} + 2*H \\
\ce{C7H4}            &  7487    &   +135.00   & \ce{C7H} + 3*H \\
\ce{CH3C6H}          &  7487    &   +135.00   & \ce{C7H4} \\
\ce{C8H}             &  6937    &    +304.5   & \ce{C7H} + C, $\Delta H_f$ from (17) \\
\ce{C8H2}            &  7387    &   +208.13   & \ce{C8H} + H \\
\ce{C8H3}            &  7837    &   +212.93   & \ce{C8H} + 2*H \\
\ce{C8H4}            &  8287    &   +200.00   & \ce{C8H} + 3*H \\
\ce{C9H}             &  7737    &   +329.11   & \ce{C8H} + C \\
\ce{C9H2}            &  8187    &   +277.49   & \ce{C9H} + H \\
\ce{C9H3}            &  8637    &   +234.08   & \ce{C9H} + 2*H \\
\ce{C9H4}            &  9087    &   +200.00   & \ce{C9H} + 3*H \\
\ce{C2O}             &  1950    &    +91.37   & \ce{CO} + C, $\Delta H_f$ from (17) \\
\ce{C3O}             &  2750    &    +75.33   & \ce{C2O} + C, $\Delta H_f$ from (17) \\
\ce{HC2O}            &  2400    &    +42.40   & \ce{C2O} + H \\
\ce{CH2CO}           &  2200    &    -20.85   & \ce{CH2} + \ce{CO} \\
\ce{HC3O}            &  3200    &    +46.23   & \ce{C3O} + H \\
\ce{c-H2C3O}         &  3650    &    +23.00   & \ce{C3O} + 2*H \\
\ce{NHNO}            &  3814    &             & \ce{HNO} + N \\
\ce{NH2NO}           &  4264    &             & \ce{NHNO} + H \\
\ce{C2N}             &  2400    &    +163.5   & \ce{CN} + C, $\Delta H_f$ from (17) \\
\ce{C3N}             &  3200    &    +173.7   & \ce{CN} + 2*C, $\Delta H_f$ from (17) \\
\ce{C4N}             &  4000    &    +210.5   & \ce{CN} + 3*C, $\Delta H_f$ from (17) \\
\ce{C5N}             &  4800    &    +231.7   & \ce{CN} + 4*C, $\Delta H_f$ from (17) \\
\ce{C7N}             &  6400    &    +287.7   & \ce{CN} + 6*C, $\Delta H_f$ from (17) \\
\ce{C9N}             &  8000    &    +354.0   & \ce{CN} + 8*C \\
\ce{HCNH}            &  2458    &    +55.70   & \ce{CH} + \ce{NH} \\
\ce{CH2NH}           &  2583    &    +26.00   & \ce{CH2} + \ce{NH} \\
\ce{CH3NH}           &  2708    &    +43.60   & \ce{CH3} + \ce{NH} \\
\ce{CH2NH2}          &  3317    &    +38.00   & \ce{CH2} + \ce{NH2} \\
\ce{CH3NH2}          &  3442    &     -5.50   & \ce{CH3} + \ce{NH2} \\
\ce{HC2N}            &  3780    &   +122.65   & \ce{CH2CN} - H, $\Delta H_f$ from (17) \\
\ce{HC3N}            &  4580    &    +84.63   & \ce{HC2N} + C, $\Delta H_f$ from (17) \\
\ce{HC5N}            &  6180    &    +140.6   & \ce{HC2N} + 3*C, $\Delta H_f$ from (17) \\
\ce{HC7N}            &  7780    &    +191.8   & \ce{HC2N} + 5*C, $\Delta H_f$ from (17) \\
\ce{HC9N}            &  9380    &    +242.9   & \ce{HC2N} + 7*C, $\Delta H_f$ from (17) \\
\ce{C2NCH}           &  4580    &    +84.60   & \ce{HC3N} \\
\ce{HC2NC}           &  4580    &    +84.60   & \ce{HC3N} \\
\ce{HNC3}            &  4580    &    +84.60   & \ce{HC3N} \\
\ce{H2C2N}           &  4230    &    +57.61   & \ce{CH3CN} - H, $\Delta H_f$ from (17) \\
\ce{CH3CN}           &  4680    &    +17.70   &  \\
\ce{HC3NH}           &  5030    &    +152.6   & \ce{HC3N} + H, $\Delta H_f$ from (17) \\
\ce{C3H3N}           &  5480    &    +42.95   & \ce{HC3N} + 2*H \\
\ce{H4C3N}           &  5930    &             & \ce{HC3N} + 3*H \\
\ce{H5C3N}           &  6380    &    +12.30   & \ce{HC3N} + 4*H \\
\ce{H2C5N}           &  6630    &    +185.3   & \ce{HC5N} + H, $\Delta H_f$ from (17) \\
\ce{H3C5N}           &  7080    &             & \ce{HC5N} + 2*H \\
\ce{H2C7N}           &  8230    &    +226.7   & \ce{HC7N} + H, $\Delta H_f$ from (17) \\
\ce{H3C7N}           &  8680    &             & \ce{HC7N} + 2*H \\
\ce{H2C9N}           &  9830    &             & \ce{HC9N} + H \\
\ce{H3C9N}           &  10280   &             & \ce{HC9N} + 2*H \\
\ce{CH3C3N}          &  6280    &    +81.00   & \ce{C3H3N} + C \\
\ce{CH3C5N}          &  7880    &             & \ce{H3C5N} + C \\
\ce{CH3C7N}          &  9480    &             & \ce{H3C7N} + C \\
\ce{NH2OH}           &  5117    &    -10.00   & \ce{NH2} + \ce{OH} \\
\ce{NH2CHO}          &  3867    &    -44.50   & \ce{NH2} + \ce{HCO} \\
\ce{NH2NH}           &  3800    &    +58.20   & \ce{NH} + \ce{NH2} \\
\ce{HNOH}            &  4383    &    +27.29   & \ce{NH} + \ce{OH} \\
\ce{NH2NH2}          &  4533    &    +22.80   & 2*\ce{NH2} \\
\ce{NH2OCH3}         &  4767    &             & \ce{NH2} + \ce{CH3O} \\
\ce{NH2CH2OH}        &  7351    &             & \ce{NH2} + \ce{CH2OH} \\
\ce{NH2CO}           &  3417    &     -3.60   & \ce{NH2} + \ce{CO} \\
\ce{HNCHO}           &  3133    &    +19.53   & \ce{NH} + \ce{HCO} \\
\ce{HNCONH}          &  4217    &             & \ce{CO} + 2*\ce{NH} \\
\ce{NH2CONH}         &  4950    &             & \ce{NH2} + \ce{CO} + \ce{NH} \\
\ce{NH2CONH2}        &  5683    &    -56.29   & 2*\ce{NH2} + \ce{CO} \\
\ce{HNCOOH}          &  6653    &             & \ce{NH} + \ce{HCOOH} - H \\
\ce{NH2COOH}         &  7387    &             & \ce{HCOOH} - H + \ce{NH2} \\
\ce{CH3ONH}          &  4033    &             & \ce{NH} + \ce{CH3O} \\
\ce{HNCH2OH}         &  6617    &             & \ce{NH} + \ce{CH2OH} \\
\ce{CH3OCONH}        &  5183    &             & \ce{CH3O} + \ce{CO} + \ce{NH} \\
\ce{HNCOCH2OH}       &  7767    &             & \ce{CH2OH} + \ce{CO} + \ce{NH} \\
\ce{CH3OCONH2}       &  5917    &   -101.60   & \ce{CH3O} + \ce{CO} + \ce{NH2} \\
\ce{NH2COCH2OH}      &  8501    &             & \ce{NH2} + \ce{CO} + \ce{CH2OH} \\
\ce{HNCOCHO}         &  4283    &             & \ce{NH} + \ce{CO} + \ce{HCO} \\
\ce{NH2COCHO}        &  5017    &             & \ce{NH2} + \ce{CO} + \ce{HCO} \\
\ce{CH3CONH}         &  3858    &             & \ce{CH3} + \ce{CO} + \ce{NH} \\
\ce{CH3CONH2}        &  4592    &    -56.96   & \ce{CH3} + \ce{CO} + \ce{NH2} \\
\ce{COOH}            &  5120    &    -43.91   & \ce{HCOOH} - H \\
\ce{HCOOH}           &  5570    &    -90.49   &  \\
\ce{CH3COOH}         &  6295    &   -103.44   & \ce{COOH} + \ce{CH3} \\
\ce{HCOCOOH}         &  6720    &             & \ce{HCO} + \ce{COOH} \\
\ce{HOCOOH}          &  7970    &             & \ce{OH} + \ce{COOH} \\
\ce{CH3OCOOH}        &  7620    &   -145.08   & \ce{CH3O} + \ce{COOH} \\
\ce{CH2OHCOOH}       &  10204   &   -139.34   & \ce{CH2OH} + \ce{COOH} \\
\ce{CH3OOCH3}        &  5000    &    -30.00   & 2*\ce{CH3O} \\
\ce{CH3OCH2OH}       &  7584    &             & \ce{CH3O} + \ce{CH2OH} \\
\ce{HOCH2CH2OH}      &  10168   &    -94.26   & 2*\ce{CH2OH} \\
\ce{CH3OOH}          &  5350    &    -31.31   & \ce{CH3O} + \ce{OH} \\
\ce{HOCH2OH}         &  7934    &    +14.79   & \ce{OH} + \ce{CH2OH} \\
\ce{CH3CO}           &  2325    &     -2.87   & \ce{CH3} + \ce{CO} \\
\ce{CH3CHO}          &  2775    &    -40.80   & \ce{CH3} + \ce{HCO} \\
\ce{CH3OCO}          &  3650    &    -39.90   & \ce{CH3O} + \ce{CO} \\
\ce{HOCH2CO}         &  6234    &    -39.20   & \ce{CH2OH} + \ce{CO} \\
\ce{HCOOCH3}         &  4100    &    -86.60   & \ce{HCO} + \ce{CH3O} \\
\ce{CH2OHCHO}        &  6684    &    -77.63   & \ce{CH2OH} + \ce{HCO} \\
\ce{CH3OCH3}         &  3675    &    -43.99   & \ce{CH3O} + \ce{CH3} \\
\ce{CH3CH2OH}        &  6259    &    -56.23   & \ce{CH3} + \ce{CH2OH} \\
\ce{CH3COCH3}        &  3500    &    -52.00   & \ce{CH3} + \ce{CO} + \ce{CH3} \\
\ce{CH3COOCH3}       &  4825    &    -98.00   & \ce{CH3} + \ce{CO} + \ce{CH3O} \\
\ce{CH3COCH2OH}      &  7409    &             & \ce{CH3} + \ce{CO} + \ce{CH2OH} \\
\ce{CH3OCOOCH3}      &  6150    &             & \ce{CH3O} + \ce{CO} + \ce{CH3O} \\
\ce{CH3OCOCH2OH}     &  8734    &   -133.10   & \ce{CH3O} + \ce{CO} + \ce{CH2OH} \\
\ce{HOCH2COCH2OH}    &  11318   &             & \ce{CH2OH} + \ce{CO} + \ce{CH2OH} \\
\ce{CH3OCOCHO}       &  5250    &             & \ce{CH3O} + \ce{CO} + \ce{HCO} \\
\ce{HOCH2COCHO}      &  7834    &             & \ce{CH2OH} + \ce{CO} + \ce{HCO} \\
\ce{COCHO}           &  2750    &             & \ce{CO} + \ce{HCO} \\
\ce{CH3COCHO}        &  3925    &    -64.80   & \ce{CH3} + \ce{CO} + \ce{HCO} \\
\ce{CHOCHO}          &  3200    &    -50.67   & 2*\ce{HCO} \\
\ce{HCOCOCHO}        &  4350    &             & \ce{HCO} + \ce{CO} + \ce{HCO} \\
\end{longtable}
\tablefoot{
\tablefoottext{a}{We include here either the literature reference to an updated value or the explicit formula based on the additive property derived from other species. In cases of no references/remarks, it is implied that the value remains the same as the previous OSU model \citep[][, or references therein]{Garrod2008}.}
} 
\tablebib{
(1)~\citet{Schuurman2004}; (2)~\citet{He2014}; (3)~\citet{Wakelam2017}; (4)~\citet{MD2014}; (5)~\citet{Jing2012}; (6)~\citet{Etim2017}; (7)~\citet{Riley2007}; (8)~\citet{Pozzo2008}; (9)~\citet{Cioslowski2000}; (10)~\citet{Steudel2003Allotropes}; (11)~\citet{Luke1985}; (12)~\citet{Griller1990}; (13)~\citet{McBride2002}; (14)~\citet{Ruscic1992}; (15)~\citet{Resende2003}; (16)~\citet{JANAF}; (17)~\citet{Etim2016}; (18)~\citet{Burcat2001}; (19)~\citet{Benson1978}.
}
}}

\section{Reaction tables}
In the tables below, we list out all the modifications introduced to the reaction network, with respect to the original OSU gas/grain astrochemical network \citep{Garrod2008}. We have separated these changes out into many different tables according to whether they pertain to sulfur or not, as well as the type of modification (removal, modification, or new additions). For the tables listing chemical reactions, we have tried to include as much information as possible to help with their adoption, including the reaction itself (i.e. the reactions and products), the kinetics that define the rate coefficient, the form of the rate law, and finally any notes/references that were used for defining the entry.

The chemical reactions include many entries that are entirely based on educated guesses, and were synthesized based on a number of general chemical/physical principles, which we describe here.

\begin{table*}\centering
\caption{Non-sulfur changes to gas/grain network.}
\label{table:nonSchanges}
\begin{tabular}{llllll}
\hline
\hline
Reaction & rtype\tablefootmark{a} & $\alpha$ & $\beta$ & $\gamma$ & Remark \\
\hline
\ce{H{}_{surf} + H2CO{}_{surf} -> HCO{}_{surf} + H2{}_{surf}} & 14 & - & - & 1740 & $E_\mathrm{a}$ from (1) \\
\ce{H{}_{surf} + H2CO{}_{surf} -> CH2OH{}_{surf}} & 14 & - & - & 5400 & $E_\mathrm{a}$ from (1) \\
\ce{H{}_{surf} + H2CO{}_{surf} -> CH3O{}_{surf}} & 14 & - & - & 2200 & $E_\mathrm{a}$ from (1) \\
\ce{H2CO{}_{surf} -> CO{}_{surf} + H2{}_{surf}} & 20 & 6.67e-11 & - & 2.8 & fixed typo in products \\
\ce{CH4{}_{surf} -> CH4} & 80 & 1e-3 & - & 1.8 & (2) \\
\ce{O{}_{surf} -> O} & 80 & 1e-4 & - & 1.8 & (2) \\
\ce{OH{}_{surf} -> OH} & 80 & 1e-3 & - & 1.8 & (2) \\
\ce{H2O{}_{surf} -> H2O} & 80 & 1e-3 & - & 1.8 & (2) \\
\ce{CO{}_{surf} -> CO} & 80 & 2.7e-3 & - & 1.8 & (2) \\
\hline
\end{tabular}
\tablefoot{
\tablefoottext{a}{The ``rtype'' of each reaction can be used to look up their respective rate coefficient using the listing shown in Appendix \ref{app:ratelaws}.}
} 
\tablebib{
(1)~\citet{Ruaud2015}; (2)~\citet{Hollenbach2009}.
}
\end{table*}

{\begin{table*}\centering
\caption{Removed sulfur reactions.}
\label{table:removedrxns}
\begin{tabular}{llllll}
\hline
\hline
Reaction & rtype\tablefootmark{a} & $\alpha$ & $\beta$ & $\gamma$ & Remark \\
\hline
\ce{S2H} + cr \ce{->} \ce{HS} + \ce{S} & 1 & 1.50e+03 & 0.0 & 0 & alternative products, (1) \\
\ce{H2} + \ce{S2H^+} \ce{->} \ce{H3S2^+} & 2 & 1.00e-14 & -1.0 & 0 & not allowed: missing second product \\
\ce{H2S} + \ce{S^+} \ce{->} \ce{H2S^+} + \ce{S} & 2 & 4.40e-11 & -0.5 & 0 & not allowed: endothermic, (2) \\
\ce{HCO^+} + \ce{SO} \ce{->} \ce{CO} + \ce{HSO^+} & 2 & 3.30e-09 & -0.5 & 0 & not allowed: endothermic, (2) \\
\ce{CO} + \ce{HS} \ce{->} \ce{H} + \ce{OCS} & 7 & 5.95e-14 & 1.12 & 8330 & not allowed: endothermic, (2) \\
\ce{HCS} + \ce{O} \ce{->} \ce{CS} + \ce{OH} & 7 & 5.00e-11 & 0.0 & 0 & negligible route, (3) \\
\ce{H2S2} + h$\nu$ \ce{->} \ce{HS} + \ce{HS} & 13 & 1.00e-09 & 0.0 & 1.7 & replaced with alt. products, (1) \\
\ce{S2H} + h$\nu$ \ce{->} \ce{HS} + \ce{S} & 13 & 1.00e-09 & 0.0 & 1.7 & replaced with alt. products, (1) \\
\ce{H2S{}_{surf}} + cr \ce{->} \ce{H2{}_{surf}} + \ce{S{}_{surf}} & 17 & 5.15e+03 & 0.0 & 0 & replaced with alt. products, (4) \\
\ce{H2S2{}_{surf}} + cr \ce{->} \ce{HS{}_{surf}} + \ce{HS{}_{surf}} & 17 & 1.50e+03 & 0.0 & 0 & replaced with alt. products, (1) \\
\ce{S2H{}_{surf}} + cr \ce{->} \ce{HS{}_{surf}} + \ce{S{}_{surf}} & 17 & 1.50e+03 & 0.0 & 0 & replaced with alt. products, (1) \\
\ce{H2S2{}_{surf}} + h$\nu$ \ce{->} \ce{HS{}_{surf}} + \ce{HS{}_{surf}} & 19 & 1.00e-09 & 0.0 & 1.7 & replaced with alt. products, (1) \\
\ce{S2H{}_{surf}} + h$\nu$ \ce{->} \ce{HS{}_{surf}} + \ce{S{}_{surf}} & 19 & 1.00e-09 & 0.0 & 1.7 & replaced with alt. products, (1) \\
\ce{H2S2{}_{surf}} + h$\nu$ \ce{->} \ce{HS{}_{surf}} + \ce{HS{}_{surf}} & 20 & 1.00e-10 & 0.0 & 2.5 & replaced with alt. products, (1) \\
\ce{S2H{}_{surf}} + h$\nu$ \ce{->} \ce{HS{}_{surf}} + \ce{S{}_{surf}} & 20 & 1.00e-10 & 0.0 & 2.5 & replaced with alt. products, (1) \\
\hline
\end{tabular}
\tablefoot{
\tablefoottext{a}{The ``rtype'' of each reaction can be used to look up their respective rate coefficient using the listing shown in Appendix \ref{app:ratelaws}.}
} 
\tablebib{
(1)~\citet{Isoniemi1999}; (2)~\citet{Vidal2017}; (3)~\citet{Loison2012}; (4)~\citet{CruzDiaz2014}.
}
\end{table*}}

{\begin{table*}\centering
\caption{Modified sulfur reactions.}
\label{table:modifiedrxns}
\begin{tabular}{llllll}
\hline
\hline
Reaction & rtype\tablefootmark{a} & $\alpha$ & $\beta$ & $\gamma$ & Remark \\
\hline
\ce{S2} + cr \ce{->} \ce{S} + \ce{S} & 1 & 1.00e+02 & 0.0 & 0 & based on (1), i.e. more stable than S2H \\
\ce{C3H2} + \ce{S^+} \ce{->} \ce{H} + \ce{HC3S^+} & 2 & 4.70e-09 & -0.45 & 0 & Table.B1.140 from (2) \\
\ce{CS} + \ce{HCO^+} \ce{->} \ce{CO} + \ce{HCS^+} & 2 & 2.40e-09 & -0.45 & 0 & (3) \\
\ce{H^+} + \ce{H2CS} \ce{->} \ce{H} + \ce{H2CS^+} & 2 & 2.50e-09 & -0.4 & 0 & Table.B1.10 from (2) \\
\ce{H^+} + \ce{H2S} \ce{->} \ce{H} + \ce{H2S^+} & 2 & 6.71e-10 & -0.5 & 0 & Table.B1.4 from (2) \\
\ce{HNC} + \ce{HS^+} \ce{->} \ce{H2CN^+} + \ce{S} & 2 & 7.70e-10 & 0.0 & 0 & Table.B1.191 from (2) \\
\ce{C} + \ce{SO} \ce{->} \ce{CO} + \ce{S} & 7 & 1.00e-10 & 0.0 & 0 & Table.B1.50 from (2) \\
\ce{C} + \ce{SO} \ce{->} \ce{CS} + \ce{O} & 7 & 1.00e-10 & 0.0 & 0 & Table.B1.50 from (2) \\
\ce{C2H} + \ce{CS} \ce{->} \ce{C3S} + \ce{H} & 7 & 2.00e-10 & 0.0 & 0 & Table.B1.81 from (2) \\
\ce{CS} + \ce{O} \ce{->} \ce{CO} + \ce{S} & 7 & 2.70e-10 & 0.0 & 760 & (4) \\
\ce{CS} + \ce{OH} \ce{->} \ce{H} + \ce{OCS} & 7 & 1.70e-10 & 0.0 & 0 & (5) \\
\ce{H} + \ce{H2S} \ce{->} \ce{H2} + \ce{HS} & 7 & 7.00e-12 & 1.7 & 455 & fit to 70-2000 K, (6) \\
\ce{HCS} + \ce{O} \ce{->} \ce{H} + \ce{OCS} & 7 & 1.00e-10 & 0.0 & 0 & Table.B1.116 from (2) \\
\ce{HS} + \ce{HS} \ce{->} \ce{H2S} + \ce{S} & 7 & 3.00e-11 & 0.0 & 2300 & (7) \\
\ce{O2} + \ce{S} \ce{->} \ce{O} + \ce{SO} & 7 & 2.10e-12 & 0.0 & 0 & (4) \\
\ce{O2} + \ce{SO} \ce{->} \ce{O} + \ce{SO2} & 7 & 1.60e-13 & 0.0 & 2300 & (4) \\
\ce{CO} + \ce{S} \ce{->} \ce{OCS} + h$\nu$ & 8 & 1.60e-17 & -1.5 & 1440 & (5) \\
\ce{e^-} + \ce{HSO^+} \ce{->} \ce{H} + \ce{SO} & 9 & 1.50e-07 & -0.5 & 0 & Table.B1.224 from (2) \\
\ce{e^-} + \ce{OCS^+} \ce{->} \ce{CO} + \ce{S} & 9 & 2.91e-07 & -0.62 & 0 & (8) \\
\ce{e^-} + \ce{OCS^+} \ce{->} \ce{CS} + \ce{O} & 9 & 4.90e-08 & -0.62 & 0 & (8) \\
\ce{OCS} + h$\nu$ \ce{->} \ce{e^-} + \ce{OCS^+} & 13 & 7.70e-10 & 2.86 & 3.85 & (9) \\
\ce{S2} + h$\nu$ \ce{->} \ce{e^-} + \ce{S2^+} & 13 & 1.30e-10 & 2.36 & 3.27 & (9) \\
\ce{S2H} + h$\nu$ \ce{->} \ce{e^-} + \ce{S2H^+} & 13 & 3.00e-10 & 2.36 & 3.27 & estimate, at least 2*S2 \\
\ce{SO} + h$\nu$ \ce{->} \ce{O} + \ce{S} & 13 & 4.20e-09 & 1.94 & 2.76 & (9) \\
\ce{SO2} + h$\nu$ \ce{->} \ce{O} + \ce{SO} & 13 & 2.40e-09 & 1.94 & 2.78 & (9) \\
\ce{H{}_{surf}} + \ce{H2S{}_{surf}} \ce{->} \ce{H2{}_{surf}} + \ce{HS{}_{surf}} & 14 & 1.00e+00 & 0.0 & 1530 & E$_\text{a}$ from (10) \\
\ce{H{}_{surf}} + \ce{OCS{}_{surf}} \ce{->} \ce{CO{}_{surf}} + \ce{HS{}_{surf}} & 14 & 1.00e+00 & 0.0 & 1950 & E$_\text{a}$ updated to match gas-phase \\
\ce{S2{}_{surf}} + cr \ce{->} \ce{S{}_{surf}} + \ce{S{}_{surf}} & 17 & 1.00e+02 & 0.0 & 0 & same as rtype=1 \\
\ce{OCS{}_{surf}} + cr \ce{->} \ce{CO{}_{surf}} + \ce{S{}_{surf}} & 18 & 1.20e+03 & 0.0 & 0 & (11) \\
\ce{OCS{}_{surf}} + cr \ce{->} \ce{CS{}_{surf}} + \ce{O{}_{surf}} & 18 & 2.40e+02 & 0.0 & 0 & (11) \\
\ce{H2S{}_{surf}} + h$\nu$ \ce{->} \ce{H{}_{surf}} + \ce{HS{}_{surf}} & 19 & 6.50e-09 & 1.8 & 2.63 & (12) \\
\ce{S2{}_{surf}} + h$\nu$ \ce{->} \ce{S{}_{surf}} + \ce{S{}_{surf}} & 19 & 6.60e-10 & 1.28 & 1.9 & (9) \\
\ce{SO{}_{surf}} + h$\nu$ \ce{->} \ce{O{}_{surf}} + \ce{S{}_{surf}} & 19 & 4.20e-09 & 1.94 & 2.76 & (9) \\
\ce{SO2{}_{surf}} + h$\nu$ \ce{->} \ce{O{}_{surf}} + \ce{SO{}_{surf}} & 19 & 2.40e-09 & 1.94 & 2.78 & (9) \\
\ce{H2S2{}_{surf}} + h$\nu$ \ce{->} \ce{H{}_{surf}} + \ce{S2H{}_{surf}} & 20 & 1.10e-10 & 0.0 & 2.5 & (1) \\
\ce{OCS{}_{surf}} + h$\nu$ \ce{->} \ce{CO{}_{surf}} + \ce{S{}_{surf}} & 20 & 2.00e-10 & 0.0 & 2.71 & (11) \\
\ce{OCS{}_{surf}} + h$\nu$ \ce{->} \ce{CS{}_{surf}} + \ce{O{}_{surf}} & 20 & 3.70e-11 & 0.0 & 2.71 & (11) \\
\ce{S2{}_{surf}} + h$\nu$ \ce{->} \ce{S{}_{surf}} + \ce{S{}_{surf}} & 20 & 7.00e-11 & 0.0 & 2.5 & (1) \\
\ce{S2H{}_{surf}} + h$\nu$ \ce{->} \ce{H{}_{surf}} + \ce{S2{}_{surf}} & 20 & 1.40e-11 & 0.0 & 2.5 & (1) \\
\hline
\end{tabular}
\tablefoot{
\tablefoottext{a}{The ``rtype'' of each reaction can be used to look up their respective rate coefficient using the listing shown in Appendix \ref{app:ratelaws}.}
} 
\tablebib{
(1)~\citet{Isoniemi1999}; (2)~\citet{Vidal2017}; (3)~\citet{Gronowski2014}; (4)~\citet{Atkinson2004}; (5)~\citet{Loison2012}; (6)~\citet{Peng1999}; (7)~\citet{Zhou2008}; (8)~\citet{Montaigne2005}; (9)~\citet{Heays2017}; (10)~\citet{Lamberts2017}; (11)~\citet{Ferrante2008}; (12)~\citet{CruzDiaz2014}.
}
\end{table*}}

{\begin{table*}\centering
\caption{New sulfur reactions.\tablefootmark{a}}
\label{table:newrxns}
\begin{tabular}{llllll}
Reaction & rtype\tablefootmark{b} & $\alpha$ & $\beta$ & $\gamma$ & Remark \\
\hline
\ce{Cl^+} + \ce{grain^-} \ce{->} \ce{grain} + \ce{Cl{}_{surf}} & 0 & 2.90e-17 & 0.5 & 0 & mass-scaled from Fe \\
\ce{Fe^+} + \ce{grain^-} \ce{->} \ce{grain} + \ce{Fe{}_{surf}} & 0 & 2.30e-17 & 0.5 & 0 & duplicated from neutral atom \\
\ce{grain^-} + \ce{Mg^+} \ce{->} \ce{grain} + \ce{Mg{}_{surf}} & 0 & 3.70e-17 & 0.5 & 0 & duplicated from neutral atom \\
\ce{grain^-} + \ce{Na^+} \ce{->} \ce{grain} + \ce{Na{}_{surf}} & 0 & 3.60e-17 & 0.5 & 0 & duplicated from neutral atom \\
\ce{grain^-} + \ce{P^+} \ce{->} \ce{grain} + \ce{P{}_{surf}} & 0 & 3.10e-17 & 0.5 & 0 & mass-scaled from Fe \\
\ce{grain^-} + \ce{S^+} \ce{->} \ce{grain} + \ce{S{}_{surf}} & 0 & 3.00e-17 & 0.5 & 0 & duplicated from neutral atom \\
\ce{grain^-} + \ce{Si^+} \ce{->} \ce{grain} + \ce{Si{}_{surf}} & 0 & 3.30e-17 & 0.5 & 0 & duplicated from neutral atom \\
\ce{CS2} + cr \ce{->} \ce{CS} + \ce{S} & 1 & 1.70e+03 & 0.0 & 0 & copied from CO2 \\
\ce{CSSH} + cr \ce{->} \ce{CS2} + \ce{H} & 1 & 1.70e+03 & 0.0 & 0 & duplicated from CS2 \\
\ce{HC2S} + cr \ce{->} \ce{CH} + \ce{CS} & 1 & 1.50e+03 & 0.0 & 0 & duplicated from HC2O \\
\hline
\end{tabular}
\tablefoot{
\tablefoottext{a}{Due to technical issues, this table has been limited to the first 10 entries. Please see the ancillary files for a full version of the PDF, as well as ASCII files containing the reaction network.}
\tablefoottext{b}{The ``rtype'' of each reaction can be used to look up their respective rate coefficient using the listing shown in Appendix \ref{app:ratelaws}.}
} 
\tablebib{(1)~\citet{Isoniemi1999}; (2)~\citet{JE2012}; (3)~\citet{Vidal2017}; (4)~\citet{Savage2004}; (5)~\citet{Shannon2014}; (6)~\citet{Loison2012}; (7)~\citet{Majumdar2016}; (8)~\citet{Atkinson2004}; (9)~\citet{Glarborg2014}; (10)~\citet{Sendt2002}; (11)~\citet{Andreazza2005}; (12)~\citet{Montaigne2005}; (13)~\citet{Heays2017}; (14)~\citet{Garozzo2010}; (15)~\citet{Yamada2002}; (16)~\citet{Ferrante2008}; (17)~\citet{Zhou2013}; (18)~\citet{Cheng1996}; (19)~\citet{Zhou2009}; (20)~\citet{JE2014}; (21)~\citet{Moore2007}; (22)~\citet{Ballester2008}; (23)~\citet{Glarborg2013}; (24)~\citet{Steudel2003Allotropes}; (25)~\citet{Sheraton1981}; (26)~\citet{CruzDiaz2014}; (27)~\citet{Hollenbach2009}; (28)~\citet{Fuente2017}.
}
\end{table*}}

\subsection{Rate laws} \label{app:ratelaws}
Following the conventions established already decades ago, the reaction network is made up of classes (or types) of reactions that can be categorized by either the types of reactants/products involved or the forms of the rate laws that define their individual reaction rates. In the reaction tables below (Tables \ref{table:nonSchanges}--\ref{table:newrxns}), we have sorted the reactions according to the reaction types, which we designate as their ``rtype''. The list below can be used to look up the rate laws that define their rate coefficients, $k_x$, according to their ``rtype'' (in subscript):
\begin{itemize}
	\item $k_0 = \alpha \left(\nicefrac{T}{300}\right)^\beta \quad (\mathrm{cm}^3 ~\mathrm{s}^{-1})$
	\item $k_{1,17--18} = \alpha \zeta \quad (\mathrm{s}^{-1})$
	\item $k_{2-12} = \alpha \left(\nicefrac{T}{300}\right)^\beta e^{-\left(\nicefrac{\gamma}{T}\right)} \quad (\mathrm{cm}^3 ~\mathrm{s}^{-1})$
	\item $k_{13,19-20,80} = \alpha e^{-\gamma A_\mathrm{v}} \quad (\mathrm{s}^{-1})$ (also see App. \ref{app:photo} below)
	\item $k_{14}$: (see App. \ref{app:grates} below)
	\item $k_{15} = \alpha \nu_0 e^{-\left(\nicefrac{E_\mathrm{bind}}{T_\mathrm{d}}\right)} \quad (\mathrm{s}^{-1})$
	\item $k_{16} = \alpha \nu_0 t_\mathrm{CR} R_\mathrm{CRFE} e^{-\left(\nicefrac{E_\mathrm{bind}}{T_\mathrm{S}}\right)} \quad (\mathrm{s}^{-1})$
	\item $k_{99} = \sigma_\mathrm{d} \left<v_i\right> n_\mathrm{d} \quad (\mathrm{s}^{-1})$
\end{itemize}

\subsection{Gas-phase kinetics: electrostatics} \label{app:estatics}
Estimates for gas-phase rates involving electrostatics are based on two general assumptions, depending on the specific case.

For the adsorption of atomic cations onto negatively-charged grains, we have simply copied the rates previously defined for the collisional rates involving the neutralization reactions of the atomic ions and negative grains, and this was done for all the heavy atoms (Fe, Mg, Na, S, Si, Cl, and P) with relatively large binding energies, in agreement with the study by \citet{UmebayashiNakano1980}.

On the other hand, capture rates between molecular ions and neutral molecules have been calculated explicitly for each case that is lacking an external reference. Capture rates have been estimated using the formulae generalized by \citet{Troe1985}, where we include both the classic Langevin rate constant, k$_\text{L}$, as well as the dipole-enhanced rate constant, k$_\text{D}$. The Langevin rate constant is defined by:
\begin{equation}
k_L = 2 \pi q \sqrt{\frac{\alpha}{\mu}},
\end{equation}
where $q$ is the charge of the ion, $\alpha$ is the isotropic polarizability of the neutral molecule, and $\mu$ is the center of mass of the two species. The dipole-enhanced rate constant is defined by:
\begin{equation}
k_D = 2 \pi q \mu_D \sqrt{\frac{2}{\pi \mu k T}},
\end{equation}
where the permanent dipole moment $\mu_D$ of the neutral species may also contribute to an enhanced, temperature-dependent collision rate, with $k$ referring to the Boltzmann constant, and $T$ being the kinetic temperature. An additional factor $C$ that is less than unity must be used to compensate for the imperfectly-aligned dipole of a rotating gas-phase molecule:
\begin{equation}
k = k_L + C k_D.
\end{equation}
At the most rigorous level of this theory, one might consider how factors such as temperature, rotation constants, internal state populations, and partition functions might all affect the factor $C$. In practice, it appears that this value ranges somewhere between ca. 0.386 and 1, and only nearer to unity for very light molecules (i.e. not S-bearing molecules) at very low temperatures ($\leq 10~K$). We therefore just use a value of $C = 0.4$. The determination of $k_L$ and $k_D$ may easily be estimated from only a few molecular properties, most of which are readily available from external references or easily calculable with computational chemistry packages. We have used Psi4 (\url{http://www.psicode.org/}) for the estimation of all proton affinities, the online service Chemicalize (\url{https://chemicalize.com/}) for estimates of molecular polarizabilities, and, when possible, the CDMS and JPL databases for molecular spectroscopy (\url{https://www.astro.uni-koeln.de/cdms/catalog} and \url{https://spec.jpl.nasa.gov/}, respectively) for dipole moments, otherwise our own simple Psi4 calculations. The uncertainties for these calculations are expected to be similar to the already-present entries: $\pm 50\%$.

The specific selection of which ion-neutral reactions to include is roughly based on which of those were already present for the O-substituted species, with the additional qualification that proton exchange reactions (i.e. $\ce{A(H)+ + B -> A + B(H+)}$) are allowed only when the relative proton affinities of the reacting species yields an exothermic reaction. Besides this manual selection of reactions that were included, the inclusion was also extended through a number of built-in routines of our astrochemical code to ensure that all new neutral species reacted with a minimal set of highly-abundant ions, and that all new ionic species could be neutralized through recombination.

\subsection{Gas-phase kinetics: neutral-neutral rates} \label{app:neutrates}
There are many reactions that satisfy the basic requirements for inclusion in our model (endothermic, none/low activation barrier, at least two products so one may serve as a kinetic/energetic sink), but have not been suitably targeted by theory or experiment to provide accurate gas-phase kinetics. Due to the fact that gas-phase rates for neutral-neutral reactions are notoriously difficult to estimate accurately, we have not attempted to include such reactions. In many of these cases, we have already included notes in the text above as suggestions for future studies.

\subsection{Photochemistry} \label{app:photo}
Not all molecular species that were addressed in this study were subjected to previous quantitative laboratory/theoretical photochemical studies. However, it is critical that all species are able to interact with cosmic-rays and/or energetic photons, and this was ensured through a built-in routine that checks for certain inadequacies within the chemical network. Therefore, we had to make a number of guesses on reaction channels and cross sections of cosmic-rays and/or UV photons. As a general rule of thumb, all the molecules considered in this study were at least addressed either directly or indirectly by particle/photon bombardment studies, and therefore relative cross sections of known species could at least be estimated.

We should also note that a number of photodissociation and photoionization reactions were updated according to the suggestions by \citet{Heays2017}. In those cases, we have also included the reaction coefficient with which the dust-attenuation factor may be calculated using a 2nd-order exponential integral in place of the standard exponential function, resulting in a rate law instead equal to: $$k_{13,19} = \alpha E_2(-\beta A_\mathrm{v}) \quad (\mathrm{s}^{-1}).$$ This alternative equation improves the fitted rates across a wider range of magnitudes and was found to be advantageous for our multi-stage model.

\subsection{Grain reaction rates} \label{app:grates}
The rates for surface reactions are computed in the same manner as that of \citet{Garrod2008}, including allowing for tunneling to be considered in case the reaction involves the light species \ce{H} or \ce{H2} and an activation barrier. In general terms, the thermally-controlled reaction rate is defined under the scheme presented in \citet{Hasegawa1992}. That is, each species has a characteristic diffusion rate based on its mass, the grain temperature, and diffusion- and binding-energies, and the reaction rates proceed according to these classical diffusion rates, activation barriers or endothermic reactions excepted. Endothermic reactions are strictly prohibited in our model, and there is logic coded in to check against all the surface reactions, both old and new. Indeed there are some cases where our tabulated binding energies and heats of formation combine to yield an endothermic reaction, and make note of this in the tables below for such cases.

\section{Details of chemistry}

\subsection{Sulfur allotropes and polysulfanes}

\subsubsection{Pure S}

One of the most significant aspects of sulfur in the condensed phase is that it exhibits the most allotropes (i.e. different forms of a pure element) known to exist for any single element in the periodic table \citep[e.g.][]{Steudel2003Allotropes}. Whereas discussion of heavy sulfur allotropes is typically limited to ambient pressures and relatively high temperature ($\gtrsim 300$ K), there have been clues supporting their importance in astrophysically-relevant ices \citep{MC2002, JE2011}. There have been instances where such chemistry was considered within the context of astrochemical modeling \citep[e.g.][]{Wakelam2005, Woods2015, Vidal2017}, but not yet in an explicit and systematic manner.

For all species \ce{S_x}, we have treated them as normal gas/grain species like all others present in the model. The diradical chains (x = 2--4) may form through barrierless atom-addition reactions on the grain, and the rings (x = 5--8) may form only through combination reactions of the smaller chains (i.e. diradical-diradical ring closures). Their binding energies are assumed to scale with the number of sulfur atoms, and therefore these aforementioned grain processes are not likely to be efficient at dust temperatures of 10--15 K. Unfortunately there is also a lack of information regarding photochemistry cross sections. Therefore, we have made the assumption that the photostabilities increases with size of the chains/rings (i.e. the cross sections decrease).

The only species within this category that reaches an appreciable gas-phase abundance is \ce{S2}, reaching a fractional abundance $X(\ce{S2}) \approx 10^{-9}$ after 0.2 Myr during the translucent cloud stage. Its abundance is nearly constant after this, including during the dense stage, except for a lull at intermediate times under dense conditions, when its formation cannot keep up with its rapid destruction by C and O atoms. On the grain, \ce{S3} reaches the highest abundance among the pure allotropes, to a fractional abundance of $\sim 10^{-11}$ after 0.2 Myr during the translucent cloud stage. Its rotational spectrum is relatively well-studied \citep{Thorwirth2005}, but its detection would prove challenging at such a low abundance. We had hoped to see more significant abundances of these pure allotropes, but it appears that even the small chains are too heavy and/or too sticky (high binding energies/diffusion barriers) for thermal roaming to react with each other before alternative processes destroy them.

\subsubsection{(Poly)sulfanes (\ce{S_xH_y})}\label{sec:sulfanes}

Aside from the pure allotropes of sulfur, a number of hydrogenated forms may also readily co-exist and are collectively known as sulfanes or polysulfanes. The simplest species of this class are \ce{HS} and \ce{HS+}, but it was only relatively recently that they were detected in the ISM \citep[][respectively]{Neufeld2012, Menten2011}, specifically within diffuse interstellar clouds, at fractional abundances of $\leq 4 \times 10^{-9}$. Our model does not predict an agreeable abundance of \ce{HS} during the diffuse stage in comparison with \citet{Neufeld2015}, however our model does reach $X(HS) \approx 10^{-9}$ late into the translucent cloud stage and it remains there for much of the dense cloud stage. On the other hand, \ce{HS+} is not predicted at a significant abundance at all during our model's early stages. \citet{Savage2004} have suggested two possible formation routes for \ce{HS+}:
\begin{ceqn}
\begin{align*}
\ce{S+ + H2} &\ce{-> HS+ + H},\\
\ce{H2S+ + H} &\ce{-> HS+ + H2},
\end{align*}
\end{ceqn}
however, the former requires an activation barrier of nearly 10000 K \citep{Millar1987}, and the latter relies on atomic hydrogen and \ce{H2S+}, neither of which are highly abundant during the dense stage and this may explain the low prediction of \ce{HS+}. This shortcoming of our cold, dense model lends support to the hypothesis that shock chemistry may play a crucial role to form these species \citep{Neufeld2015}.

\ce{H2S} becomes an important molecule in the gas-phase during the translucent cloud stage and for much of the dense cloud stage, with a near-constant fractional abundance of $\sim10^{-9}$. In the gas-phase, the main formation route of \ce{H2S} during the translucent cloud stage (i.e. low-$A_v$) is photodesorption of its grain component, which was missing completely from previous astrochemical models, and then during the dense stage from \ce{H2S+} via non-dissociative recombination with electrons and charge-transfer reactions involving a slew of other heavy atoms. On the grain, \ce{H2S} is formed rapidly through successive hydrogenation of elemental sulfur and is therefore an important primary constituent of sulfur on the grain. During the early stages of sulfur depletion, it begins to account for up to ca. 10\% of the total sulfur budget, it never builds up to more significant levels, as cosmic rays keep its abundance in check while a number of recently-adsorbed species---notably atomic carbon---scavenge it and convert it to other species. The limited peak abundance of solid \ce{H2S} agrees very well with interstellar non-detections, but stands in stark contrast with the cometary value reported by \citet{Calmonte2016}, who suggest that it accounts for the majority of the sulfur composition in the bulk ice of comet 67P/Churyumov-Gerasimenko. Additionally, we would like to point out that very recent results \citep{Oba2018} suggest that \ce{H2S} formation in ice may also have a non-negligible reactive desorption, which may drive down its abundance on the grain even lower than our model predicts.

\ce{H2S+} itself can be formed quite efficiently from the radiative association reaction, $$\ce{S+ + H2 -> H2S+} + h\nu,$$ which was already included in the OSU network, as well as $$\ce{H3+ + S -> H + H2S+}.$$ The reaction above was introduced by \citet{Vidal2017} as a second route with a relatively small branching ratio in comparison to the route that yields the products \ce{H2}/\ce{HS+}. On the grain, \ce{H2S} may form rapidly through association of the \ce{HS} radical with atomic hydrogen. The ionic (\ce{H2S+}) and protonated (\ce{H3S+}) forms are also predicted to reach non-negligible gas-phase abundances ($\sim 10^{-11}$) at late-times during the dense stage. These ions may prove fruitful targets for astronomical detection, providing better constraints for astrochemical modeling, and in the case of \ce{H2S+}, a good target for additional laboratory spectroscopy.

Two other important (poly)sulfanes are \ce{S2H} and \ce{H2S2} (also sometimes written as \ce{HSSH}), the former of which has now recently been reported in the interstellar medium \citep{Fuente2017}. These species have long been suspected as interstellar candidates, because of the importance of \ce{H2S} and how readily it may be converted to alternative chemistry upon irradiation \citep{JE2011}. Whereas our model predicts a near-constant abundance of \ce{S2} after a specific time during the translucent cloud stage, our model predicts only a significant gas-phase abundance of \ce{S2H} and \ce{H2S2} during the dense stage, and their abundances are predicted to nearly match each other at a peak fractional abundance of $\sim10^{-10}$, which is comparable the detection reported by \citet{Fuente2017} toward the Horsehead PDR. \ce{S2H} always reaches a higher peak abundance in the gas-phase compared to the grain, as it lacks an efficient formation route on the grain, however \ce{H2S2} does reach a higher abundance in the ice by a factor of 10. In the gas-phase, \ce{S2H} and \ce{H2S2} are expected to form from a similar but rather convoluted ionic pathway:
\begin{ceqn}
\begin{align*}
\ce{H3+ + S2} &\ce{-> S2H+ + H2},\\
\ce{S2H+ + H2} &\ce{-> H3S2+ + h\nu},\\
\ce{H3S2+ + e-} &\ce{-> S2H + H2}~(50\%),\\
 &\ce{-> H2S2 + H}~(50\%),
\end{align*}
\end{ceqn}
where the final dissociative recombination reaction is estimated to have a branching ratio of 1:1 between the two sets of products. This final reaction originates from the OSU gas/grain model, but we could not find any experimental or theoretical studies to support this particular branching ratio; any update to this product branching ratio would have a significant impact on the relative abundances of \ce{S2H} and \ce{H2S2} in the gas phase. Under our model conditions, the ions involved in the aforementioned pathway are not predicted at high abundance, but they aren't well-studied in the laboratory and knowledge of their rotational spectra might yield some surprises under certain interstellar conditions. On the grain, \ce{S2H} has no significant formation pathway except for the photo- and CR-induced dissociation reactions of \ce{H2S2}. \ce{H2S2} primarily forms on the grain from the reaction, $$\ce{S + H2S -> H2S2},$$ which has been shown to be barrierless despite involving a closed-shell species and a number of transient intermediates \citep{Zhou2008}. However, the reaction itself is not particularly efficient because of the reactants' high surface binding energies (ca. 1100 K and 2700 K, respectively), and it benefits only from their high abundances. \ce{H2S2} may also form from the reaction of the \ce{HS} radical with itself, but this reaction must compete with the more rapid hydrogenation of \ce{HS} to form \ce{H2S}.


\subsection{C-bearing species}

\subsubsection{\ce{CS}}

\ce{CS} is one of the most abundant S-bearing molecules under all physical conditions considered. It has significant abundances in translucent and dense stages, and is always the most abundant C-bearing sulfur species in our model. During the translucent and dense stages, our model predicts a fractional abundance of $10^{-8}$--$10^{-7}$ at intermediate times, which then drops at later times, and its abundance at these later times agrees well with observations. On the grain, its abundance during the translucent stage is similar to the gas phase, but it is enhanced an additional factor of 10 during the dense stages. Its gas-phase formation is primarily ($\sim80$\% during translucent stage, $\leq30$\% during the dense stage) through the series of ion-neutral reactions:
\begin{ceqn}
\begin{align*}
\ce{CH + S+} &\ce{-> CS+ + H},\\
\ce{CS+ + H2} &\ce{-> H + HCS+},\\
\ce{HCS+ + e-} &\ce{-> CS + H}.
\end{align*}
\end{ceqn}
In the dense stage, its gas-phase formation is also enhanced by atomic carbon stripping sulfur from \ce{SO} and \ce{OCS}.

Its grain-surface formation is largely driven by the reaction between atomic \ce{C} and the \ce{HS} radical, however its abundance also significantly relies on the accretion of gas-phase \ce{CS}, as well as radical-radical reactions that result in fragmentation. The \ce{CS+} cation is predicted to reach a fractional abundance of $\sim10^{-11}$ during intermediate times of the translucent cloud stage, however its lower dipole moment and much lower abundance compared to its neutral counterpart would likely hamper an astronomical detection.

\subsubsection{\ce{HCS}, \ce{H2CS}}

At high densities, \ce{HCS} and \ce{H2CS} are also abundant C-bearing sulfur species in the ice, and the latter is also present at a significant level in the gas-phase, reaching peak fractional abundances $\geq 10^{-9}$. \ce{HCS} is formed in the ice from the well-studied \citep[see][and refs therein]{Deeyamulla2006} insertion reaction between atomic \ce{C} and \ce{H2S}. \ce{HCS} may also reach reasonably high abundances in the gas-phase at intermediate densities ($\sim10^4$), via two neutral-neutral reactions:
\begin{ceqn}
\begin{align*}
\ce{CH2 + S} &\ce{-> H + HCS},\\
\ce{C + H2S} &\ce{-> H + HCS}.
\end{align*}
\end{ceqn}
The recent detection of \ce{HCS} toward L483 at a fractional abundance of $2 \times 10^{-10}$ \citep{Agundez2018} agrees perfectly with our dense stage model at a gas-phase density of $10^4$ cm$^{-3}$. On the grain, \ce{H2CS} relies significantly on the availability of \ce{HCS}, whereby \ce{H2CS} may form directly from the hydrogenation of \ce{HCS}. In the gas-phase, it may derive from two possible routes: the ion-neutral/DR reaction combo that is important while the ionization fraction remains high:
\begin{ceqn}
\begin{align*}
\ce{CH4 + S+} &\ce{-> H + H3CS+},\\
\ce{H3CS+ + e-} &\ce{-> H + H2CS}.
\end{align*}
\end{ceqn}
At late timescales, it may also form directly from neutral-neutral displacement reaction, $$\ce{CH3 + S -> H + H2CS}.$$

\subsubsection{Methyl mercaptan (\ce{CH3SH})}

Methyl mercaptan is the S-bearing analog of methanol (\ce{CH3OH}), and as such, much of its chemistry is similar in our model. In the gas-phase, \ce{CH3SH} is severely lacking in abundance in our dark cloud model, with a peak abundance a factor of nearly $10^5$ below that of \ce{CH3OH}. Its most efficient gas-phase formation route in the model is through the ion-neutral radiative association reaction, $$\ce{CH3+ + H2S -> CH3SH2+ + h\nu},$$ followed by successive dissociative recombination, $$\ce{CH3SH2+ + e- -> CH3SH + H}.$$ This analogous route for \ce{CH3OH} is much less efficient than the ion-neutral reaction, $$\ce{OH- + CH3 -> CH3OH + e-}.$$ However the analogous sulfur reaction of the latter route, as well as the species \ce{HS-}, are both missing from our model due to uncertainties behind them.

In the ice, on the other hand, our model shows that \ce{CH3SH} can be efficiently formed already during the translucent stage, at a fractional abundance $>10^{-10}$; and in the most dense trials, up to $10^{-7}$. Its grain surface formation is synonymous with that of \ce{CH3OH}: it likely forms through successive hydrogenation of the \ce{CH2SH} and \ce{CH3S} radicals. These precursor radicals are also synonymous with the O-bearing counterparts, whereby they may form from hydrogenation of \ce{H2CS} through high barriers \citep[800 K and 1200 K to form \ce{CH2SH} and \ce{CH3S}, respectively,][]{Vidal2017}, or from the barrierless radical-radical reactions,
\begin{ceqn}
\begin{align*}
\ce{CH2 + HS} &\ce{-> CH2SH},\\
\ce{CH3 + S} &\ce{-> CH3S}.
\end{align*}
\end{ceqn}
Our predicted peak abundances of \ce{CH3SH} in the ice are only slightly higher than its gas-phase detections toward dense protostellar environments \citep[ca. $10^{-10}-10^{-8}$,][]{Gibb2000, Mueller2016, Majumdar2016}, in agreement with previous suggestions that it is predominantly formed on the surface. Without a process to release it into the gas phase, it may not be an important component of gas toward dark clouds.

\subsubsection{Carbon-sulfur chains (\ce{H_xC_yS})}

A great variety of mixed carbon chains are known to exist in the ISM, including those containing sulfur. The OSU network was already quite developed for this class of molecules, but we have also added a significant number of reactions. As can be seen in the plots within Figure \ref{fig:stage3_sobs_dense}, our model predictions for \ce{C2S} are in good agreement with observations during the dense cloud stage at lower densities. On the other hand, \ce{C2S} appears to be quite sensitive to density, dropping off quickly and below the plotted range at higher densities.

In the gas phase, \ce{C2S} is formed primarily via the somewhat convoluted ionic pathway,
\begin{ceqn}
\begin{align*}
\ce{S+ + C2H} &\ce{-> C2S+ + H},\\
\ce{C2S+ + H2} &\ce{-> HC2S+ + H},\\
\ce{HC2S+ + e-} &\ce{-> C2S + H}.
\end{align*}
\end{ceqn}
None of these intermediate cations reach appreciable abundances. On the grain, \ce{C2S} mostly forms via the neutral-neutral reaction, $$\ce{CH + CS -> H + C2S},$$ benefiting from the high abundances of \ce{CH} and \ce{CS}.

Also noteworthy is the fact that our model predicts a significant amount of gas-phase \ce{HC2S} during the dense stage, even more than \ce{C2S}. This result is similar to the report by \citet{Agundez2015} and their observation of more \ce{HC2O} than \ce{C2O}. During the dense stage, \ce{HC2S} is formed solely through the reaction \ce{C + H2CS -> H + HC2S}; \citet{Yamada2002} suggests a number of other potential neutral-neutral gas-phase routes that may enhance its abundance even further, however the kinetics of these additional reactions remain so highly uncertain that we have left them out of the gas-phase portion of the network.

Larger chains have somewhat different formation routes, relying heavily on rapid neutral-neutral reaction even in the gas phase, in line with the investigations reported by \citet{Yamada2002} and references therein. The gas-phase formation of \ce{C3S} is partially driven by a similar ionic pathway as that of \ce{C2S}, however approximately half its formation is also due to contributions from the following neutral-neutral reactions during the dense stage:
\begin{ceqn}
\begin{align*}
\ce{C + HC2S} &\ce{-> H + C3S},\\
\ce{C2H + CS} &\ce{-> C3S + H},\\
\ce{S + C3H} &\ce{-> C3S + H}.
\end{align*}
\end{ceqn}
On the grain, only the reaction $$\ce{C + C2S -> C3S}$$ dominates. \ce{C4S} forms in a much simpler way, whereby the reaction $$\ce{S + C4H -> C4S + H}$$ accounts for its formation in both the gas phase and on the grain, though this process may not be efficient enough to provide an easy interstellar detection. There is laboratory data for the larger chains \citep[i.e. $\geq 5$ carbon atoms,][]{Gordon2001, Gordon2002}, but we have left them out of our network; both our model and observations show that the gas-phase abundance rapidly drops off for larger chains.

\subsubsection{Carbon disulfide (\ce{CS2})}

There has been much interest in \ce{CS2} raised by laboratory studies that have observed its formation within irradiated ice mixtures intended to simulate interstellar processes \citep[e.g., ][]{Ferrante2008, Garozzo2010}, and we have therefore introduced this species into the network. In the gas-phase, its presence would be difficult to detect, as its structure lacks a permanent dipole moment. In the condensed phase, it reaches a non-negligible fractional abundance at late times during the translucent cloud stage ($>10^{-10}$), and becomes a significant sink of elemental sulfur at late times of the dense stage. It can form from the following reactions,
\begin{ceqn}
\begin{align*}
\ce{C + S2} &\ce{-> CS2},\\
\ce{S + CS} &\ce{-> CS2},\\
\ce{S + HCS} &\ce{-> CS2 + H},
\end{align*}
\end{ceqn}
where the first entry is important only briefly during early times of the dense stage, and the latter two routes account for the majority of its formation, at late times of both the translucent and dense cloud stages.

\subsubsection{\ce{HCSSH}}

We have also introduced the molecule dithioformic acid (\ce{HCSSH}) to our gas/grain network, due to its focus in a recent laboratory study \citep{Prudenzano2018}. Although not much is known about this species in terms of kinetics/thermodynamics, its similarity to formic acid (\ce{HCOOH}) provides a number of insights that may be used for generating a minimal reaction network.

In the gas phase, it may be formed either directly through the neutral-neutral reaction, $$\ce{HS + H2CS -> HCSSH + H},$$ in a similar way to that of HCOOH, or also the ion-neutral reaction, $$\ce{HCSSH2+ + e- -> HCSSH + H}.$$ The latter pathway is much less important for \ce{HCSSH} than it is for \ce{HCOOH}, as its protonated form does not reach as high abundance as its O-containing counterpart. On the grain, it may form from a series of neutral-neutral addition reactions via radicals:
\begin{ceqn}
\begin{align*}
\ce{HS + CS} &\ce{-> CS2H},\\
\ce{H + CS2H} &\ce{-> CS2H}.
\end{align*}
\end{ceqn}

Its grain surface formation appears to be more efficient than the gas-phase route, with its gas-phase fractional abundance reaching ca. $10^{-13}$ at the moment when other sulfur species peak, and its grain abundance being mostly constant at $10^{-12}$ throughout the dense stages. Despite its similarity to \ce{HCOOH}, \ce{HCSSH} does not seem to be as important for sulfur chemistry as \ce{HCOOH} is for oxygen chemistry.


\subsection{N-bearing species}

\subsubsection{\ce{NS} and ``friends''}

The \ce{NS} radical is the most important interstellar nitrogen/sulfur species, having been detected in a variety of environments, including a PDR \citep{Leurini2006}, a molecular outflow \citep{SanchezContreras2000} and several dense regions \citep{Gottlieb1975, McGonagle1994}. During the translucent cloud stage, it forms in the gas phase primarily via the reaction $$\ce{N + SH -> NS + H},$$ which also drives its formation on the grain surface. At higher densities and after nitrides have been established, the ionic pathway involving the reactions,
\begin{ceqn}
\begin{align*}
\ce{NH2+ + S} &\ce{-> HNS+ + H},\\
\ce{HNS+ + e-} &\ce{-> NS + H},
\end{align*}
\end{ceqn}
also becomes somewhat (up to $30\%$) important, although the former reaction is a significant bottleneck because \ce{NH2+} can much more easily react with \ce{H2}, up to $10^4$ times more rapidly. On the grain, it becomes an important sink for sulfur in our model, accounting for up to 7--10\% of the total sulfur budget.

The ubiquity of \ce{NS} suggests that other interstellar species may easily derive from it. The closest related species is the \ce{NS+} cation, which has recently been studied in the laboratory and detected in the ISM toward B1-b as reported by \citet{Cernicharo2018}. \ce{NS+} is a closed-shell ion that may form from ion-molecule reactions between \ce{NS} and a number of cations serving as electron scavengers due to sulfur contributing to a relatively low electron affinity compared to the higher-row elements. Our model predicts a marginal fractional abundance ($<10^{-13}$) during the translucent cloud stage, primarily through charge-exchange reactions with atomic cations,
\begin{ceqn}
\begin{align*}
\ce{C+ + NS} &\ce{-> NS+ + C},\\
\ce{H+ + NS} &\ce{-> NS+ + H}.
\end{align*}
\end{ceqn}
Our model predicts peak fractional abundances of \ce{NS+} in the range of ca. $10^{-12}$--$10^{-11}$, which is in very good agreement with its detection toward B1-b. We also predict that \ce{NS+} might also form directly from photoionization of \ce{NS}, since \ce{NS} has an estimated ionization potential of $\sim 8.87$ eV \citep{Dyke1977}, making it an easy target in PDRs. This latter process, however, has not yet been studied to our knowledge in the laboratory nor by theory.

\subsubsection{\ce{HNCS} / \ce{HSCN} / \ce{HCNS} isomeric family}

The isomeric family containing isothiocyanic acid (\ce{HNCS}), thiocyanic acid (\ce{HSCN}), and thiofulminic acid (\ce{HCNS}) has received some focus from the astrochemical community. Of the three isomers listed here, \ce{HNCS} is the lowest-energy species, and the first to be detected in the ISM \citep{Frerking1979}. The slightly less stable isomer \ce{HSCN} was finally detected much more recently at an abundance slightly below that of \ce{HNCS} \citep{Halfen2009}. The two least stable isomers, \ce{HCNS} and \ce{HSNC}, have not yet been reported in the ISM, though it was only very recently that their rotational spectra were characterized sufficiently to merit an interstellar search \citep{McGuire2016}.

The chemical network pertaining to these species was adapted solely from that of \citet{Vidal2017}, as these molecules have not been well-studied in the laboratory. It should also be noted that the least stable isomer, \ce{HSNC}, remains missing from the astrochemical model. Given these circumstances, there is little we can add to the understanding of these isomers, except to state that they are not well-modeled with our model. Because we do not consider reactive desorption in our model---in contrast to the work reported by \citet{Vidal2017}---this highlights some interesting behavior in the formation of these isomers. In the gas-phase, \ce{HCNS} is produced most efficiently in comparison to the other isomers, \ce{HNCS} is predicted at an abundance ca. $10x$ below \ce{HCNS}, and \ce{HSCN} is nearly another 10-fold lower. On the grain, on the other hand, \ce{HNCS} and \ce{HSCN} have identical formation routes and (estimated) surface properties, and therefore their abundances match at all times. During the dense stage, \ce{HNCS} and \ce{HSCN} are enhanced compared to \ce{HCNS} at early times (i.e. much earlier than the free-fall time), and this role is reversed at later times. It is conceivable that reactive desorption could significantly alter the gas-phase abundances of these isomers, causing behavior that reproduces that of \citet{Vidal2017}. On the other hand, the chemistry of these species are not yet well understood, and they could significantly benefit from further physico-chemical laboratory studies.


\subsection{O-bearing species}

\subsubsection{\ce{SO} and \ce{SO2}}

With the exception of the diffuse cloud stage, \ce{SO} is predicted to be the most abundant O-bearing sulfur species in the gas phase, however our model does not match many observations, being underpredicted during the early stages and overpredicted at higher densities. During the diffuse stage, its primary formation mechanism in our model is the gas-phase reaction, $$\ce{O + HS -> SO + H},$$ and its destruction is driven by photodissociation and photoionization. During the translucent cloud stage, these processes still dominate, but its destruction is now also governed by the reaction with \ce{C+}, until later times, when carbon has been sufficiently neutralized. During the dense stage, the reaction $$\ce{OH + S -> H + SO}$$ also becomes an important route. Whereas gas-phase \ce{SO} is an important S-bearing molecule, the majority of \ce{SO} can actually be found on the grain, where the aforementioned neutral-neutral reaction may also take place, as well as the simple direct addition reaction of S and O atoms. This latter process is no longer as efficient as in previous models, due to the large binding energy of the oxygen atom.

\ce{SO2} is also an abundant O-bearing sulfur species in the gas-phase, albeit only at higher densities. In general, it peaks in gas-phase abundance approximately $10x$ lower than \ce{SO}, in general agreement with most astronomical observations. It may form directly from \ce{SO}, through the radiative association reaction $$\ce{O + SO -> SO2 + h\nu}.$$ On the grain, \ce{SO2} plays a lesser role at such low temperatures, as this latter reaction is still possible but suffers from the high binding/diffusion energies of both \ce{O} and \ce{SO}. \ce{O2} is slightly more reactive, and provides a formation route via $$\ce{O2 + SO -> SO2 + O},$$ but not at significant levels.

The radical \ce{HSO} is a derivative of \ce{SO} and also an unexpectedly important oxygen-containing species in our model despite the recent report of its gas-phase non-detection toward several extraterrestrial environments \citep{Cazzoli2016}. In our model it does not reach appreciable abundance in the gas-phase, however it appears to collect significantly on the grain, as it is modeled to form readily from the hydrogenation of \ce{SO}. Its peak abundance during the dense stage and at late times accounts for 5--20\% of the sulfur budget, which is admittedly quite high for an open-shell species.

\subsubsection{Carbonyl sulfide (\ce{OCS})} \label{sec:ocs}

In the dense stage, our model consistently predicts a peak fractional abundance of $\sim 2 \times 10^{-9}$ in the gas-phase, which sits squarely within the range of values observed in interstellar environments. Its formation derives from the following reactions (in order of decreasing contribution):
\begin{ceqn}
\begin{align*}
\ce{HCS + O} &\ce{-> H + OCS}, \\
\ce{HCO + S} &\ce{-> H + OCS}, \\
\ce{CS + OH} &\ce{-> H + OCS}.
\end{align*}
\end{ceqn}
Its destruction is dominated by carbon atoms that cleave the \ce{C-O} bond, yielding \ce{CO} and \ce{CS} radicals. On the grain, it is also one of the primary sinks of elemental sulfur, forming directly from the association of \ce{CO} and sulfur atoms. Its destruction is controlled simply by photodissociation from cosmic rays, though there is non-negligible contribution from sulfur atoms scavenging to form \ce{S2} and \ce{CO} at later times, after atomic sulfur is able to deplete onto the grain.

\end{appendix}


\end{document}